# Modelling, design and control of middle-size tilt-rotor quadrotor

**Theodore Nye-Matthew, Xinhua Wang**
Aerospace Engineering
University of Nottingham, UK
Email: wangxinhua04@gmail.com

## ABSTRACT
This paper explores the mathematical modelling and 3D design of a tilt-rotor quadrotor aircraft. The aircraft is a VTOL design and has capacity for one pilot. The design incorporates a part manual part automatic computerised flight control system and hybrid powertrain providing energy to eight ducted contrarotating propellers. Analysis of controllability was performed using the parameters derived from the developed model for the take-off phase of flight. The aircraft is of a lightweight design and is intended to fill a niche in the aviation market with potential for civilian and military applications. The aircraft has multiple redundancies in the instance of propeller failure. The viability of a hybrid powerplant is explored by combining industry standard gas turbine technology with electrical motor and battery systems.  The combination of these systems results in a safe, versatile aircraft that can operate at different levels of automation depending on environmental factors or phase of flight.  Simulation confirms the design of the aircraft.

# NOMENCLATURE

| | |
|---|---|
| $CLTF$ | Closed loop transfer function |
| $dl$ | Disk loading |
| $D$ | Drag |
| $E$ | Endurance |
| $F$ | Force |
| $F(s)$ | Force (s domain) |
| $FF\_rate$ | Fuel flow rate |
| $g$ | Surface gravity |
| $G_c$ | Controller transfer function |
| $G_s$ | Transfer function of controller and model |
| $G_0$ | Model transfer function |
| $J_{motor}$ | Moment of inertia of motor components |
| $k_d$ | Derivative error gain |
| $k_f$ | Propeller constant |
| $k_i$ | Integrated error gain |
| $k_p$ | Proportional error gain |
| $K_d$ | Motor derivative error gain |
| $K_i$ | Motor intr error gain |
| $K_m$ | Motor constant (simplified) |
| $K_p$ | Motor proportional error gain |
| $K_T$ | Torque constant |
| $L_f$ | Inductance of motor |
| $L_m$ | Inductance of motor (simplified) |
| $m$ | Mass |
| $P$ | Power |
| $R$ | Range |
| $R_f$ | Resistance of motor |
| $R_m$ | Resistance of motor (simplified) |
| $RoC$ | Rate of Climb |
| $s$ | S domain |
| $T$ | Thrust |
| $V$ | Velocity |
| $V(s)$ | Voltage (s domain) |
| $x$ | Displacement |
| $\dot{x}$ | Velocity |
| $\ddot{x}$ | Acceleration |
| $X(s)$ | Displacement (s domain) |
| $\lambda_{up}$ | Air resistance during altitude gain |
| $\rho$ | Air density |
| $v$ | Airflow velocity |

# 1 Introduction

Rotorcraft or helicopters have become a well ingrained and appreciated aircraft type, as there defining characteristic, the ability to take off and land vertically without a runway, allows them to compete against conventional aircraft, due to the versatility provide by this function. This ability also allows helicopters to fill a variety of roles not able to be accomplished by fixed wing aircraft, such as; search and rescue, continuous ground support and pursuant reconnaissance. However, when directly comparing helicopters with similar fixed-wing aircraft, they display a wide variety of flaws. These include a reduction in speed, a reduced range, a reduced payload capacity, increased noise pollution and reduction in maximum altitude. All these issues are valid but fail to include the most pressing difference. The inherent failure rate, which leads to a four times more likely chance to be involved in a fatal air accident when compared to a fixed wing aircraft of the same requirement [1].

This paper is focused on is the inherent failure characteristics of the lifting surfaces of all rotors. Due to the helicopters singular rotor if anything interrupts the rotation of the rotor the helicopter loses all lift and will be unable to maintain attitude or altitude. This effect is not mitigated by an increased number of reliable engines as engine number and reliability does not affect the helicopters ability to auto rotate or glide when the rotors are jammed or stuck. Autorotation is the method where a helicopter can safely glide to ground after an engine out event. This effect can only take place while the rotors remain able to rotate, and so if they become stuck or damaged, the aircraft has no ability to recover.

The problem of increasing the safety margins of VTOL aircraft forms the basis for the paper as it became clear that the only way to mitigate the failure of the rotating lifting surface was to increase the number of lifting surfaces available, so that if one became jammed or stuck the others would be able to compensate accordingly. This led to the current paper to produce a design of a large scale multirotor intended to be a failure safe augmentation for existing helicopter fleets. The new mechanism for lift generation presented may challenges which needed to be overcome before a sound design could be produced. While the design has a larger size, some fundamental equations and operations remain the same.

In this paper, we implement:

1) The production a tilt-rotor framework and mathematical model from which others may be able to produce a working prototype
2) The design of a preliminary layout informed by key parameters
3) Suitable existing technologies, such as composite materials and hybrid powertrains, to be implemented
4) An analysis of the controllability of the design

## 1.1 Modelling
The mathematical model for a quadrotor has the most consensus, due to its formation. As the model is derived from the physical aircraft itself, very little variation from the nominal model can be achieved [2] [3] [4]. As can be seen, most papers have an agreed upon standard. However, some variation does exist between the body frame x, y and z axis body directions. While the pervious papers use the first propeller as the direction of x body direction [5], others use a variation of 45 degrees off, or called 'plus' configuration, forming a new set of equations [6]. The difference that can be inferred from this is that x body direction simplifies the model, while complication drone movement control further on, while when using 'plus configuration', a more complicated model is produced, however it will simplify pitch and roll control further on during development control system. The model selected used the 'X' configuration.

Some variation exists between these models and the current design of the quadrotor. The most effective solution was to simplify the force generated by these contrarotating props as the force from each rotor.

## 1.2 Control methods

When analysing controller design, more variation between papers exposes itself. The most concise approach was demonstrated by [6]. To achieve this, Mr. Douglas removed the coupling between pitch and roll angles and altitude, which is only required when extreme pitch and roll angles are expected. This leads to a control system suitable for modelling hovering and slow-moving quadrotors. Very useful learning the fundamentals of control, however when compared to other control architecture design, the lack of inclusion of these terms could lead to undesirable effects. [7] has a more complicated control method, whereby the position and attitude dynamics are linked. This more complicated approach will be used when analysing the control systems later in the paper, as the different characteristics caused by the failure will need to be analysed in full detail.

Lastly, we will discuss the different mechanisms for generating values of velocity and other important derivatives. The common approach to generate these derivates is by taking in sensor data and directly deriving it mathematically, and then passing on the signal to the controller, as shown by [6] and [8] using devices such as an IMU or altitude pressure sensor. This approach carries a distinct problem, that of derivative noise. When a signal is taken directly from a sensor and derived, any noise in that signal will be amplified, and without mitigation by a filter, can lead to instability. As a result, a relatively new observer approach has been taken by [9]. When compared to the previous method, the observer designed control can estimate the values mentioned prior, while naturally producing much less noise. This leads to a much less delayed signal that a derivative and filter combination, and so much more useful for rapid reaction systems, such as a quadrotor.

## 1.3 Failure effect

Another area of research relevant is that of failure tolerant systems for drones. [10] [11] [12]. Again, there is difference between the approach to compensate for the failure if actuators. While both [10] and [11] describe using linearized mathematical model to define a double control loop, [12] focuses on using integrated observer design to examine the magnitude of faults, then the remaining actuators are directed to maintain stability. The difference stems from the double control loop creating a stable rotating system in the inner loop, and a trajectory controller for the outer loop, allowing the quadrotor to maintain altitude and heading despite appearance of extreme rotation. The fault tolerant approach [13] however aims to analyses how drastic the fault is before overcompensating. Both lead to the same outcome however, of a stable rotating system.

analyses how drastic the fault is before overcompensating. Both lead to the same outcome however, of a stable rotating system.

## 1.4 Scale up

One of the most difficult challenges to overcome regarding scaling up is that of engine to propeller power precision. While this is not a consideration for the small DC motors used in lightweight drones, when using a larger motor, the response time between input and output increases significantly. As a result of this, the drone becomes much less manoeuvrable and less viable as a control method. In combination with this, larger aircraft need more power dense powerplants to fly for worthwhile periods of time [14]. This leads to the turboshaft and more energy dense fuels. These engines provide even less responsive effect and are unsuitable for direct speed control. Methods to mitigate this effect will be analysed later in the paper.

## 1.5 Actuators- Electric motors models

Electric motors play a pivotal role in simulating the response of the craft under the take-off condition. As the aircraft requires precise control torques to be produced from the propellers, accurate motor control is required in order to determine whether the motor selection is adequate for this task. As a result several papers examining the integration were chosen. [15] [16]. While [16] describes a simple gain to be used to convert the voltage controller output to the force of the motor, [15] used a transfer function to convert between voltage and force, shown by equation (1).First the voltage is taken from the controller and fed into the following transfer function.

$$\frac{K_m}{L_f \cdot s + R_f} \qquad \ldots(1)$$

The output from the above function is the motor torque, which then acts on a propeller modelled in equation (2).

$$\frac{1}{J_{motor} \cdot s + b} \qquad \ldots(2)$$

The output from the propeller is an angular velocity value which is then fed into equation (3).

$$k_f(\cdot)^2 \qquad \ldots(3)$$

This function relates the square relationship between angular velocity and force, multiplied by a constant. The output from this is Force. This force can then me multiplied by a further gain to produce a value for control torque, which is then fed into the aircraft dynamics. This combination however proved to be too complicated for the basic control methods used, and so were simplified to directly relate Voltage to Force, shown in equation (4).

$$\frac{K_m}{L_f \cdot s + R_f} \times K_T \qquad \ldots(4)$$

This function includes the required motor delay, but without the unnecessary delay of the propeller, as the angular inertia of the propeller was included in the gain of the motor.

## 2 VTOL CONCEPT CRERATION

In this section, the process for producing results from the tilt-rotor concept and the required programmes will be explained. Section 2.1 will explain the methods used to find suitable concepts to integrate into the design. Section 2.2 will describe the process of taking initial estimates and refining then using to values concurrent with existing technology. Section 2.3 will then look at the process for modelling the design on 3DExperience and the detail required from each model.

### 2.1 VTOL AND TILT-ROTOR

Initially a wide variety of previous examples were to be analysed. The analysis from these was then put into initial drawings and used to form the basic ideas from which was to be drawn the preliminary design.

The process started with a collection of a wide variety of existing vehicles, with a variety of VTOL and STOL vehicles. These include;

- The MV-22, focus on its disk loading and powerplant issues
- The X-22 with focus on its failure to become a successful prototype
- A wide range of toy quadrotors for their design layout and controllability
- The Ehang AAV air taxi for its failure safe systems such as redundant propellers
- Military helicopters such as the MH-6 as its weight is similar to that of the proposed vehicle
- The Mosquito range of ultralight helicopters for a parallel comparison, and estimates of characteristics

After collection of the vehicles the design characteristics were scrutinised to find which would be useful in the concept of the small tilt rotor, and which would be lost. For example, the ducted propellers of the X-22 were found to be efficient in forward flight but being directly driven by four individual engines lead to several accidents and crashes, as the propellers themselves were controlled rather than the engines. As a result, the use of ducted propellers was kept, but the reliance on individual propeller control would be lost.

### 2.2 Parameter selection for model and control is considered

After considering the basic structures and characteristics to be included, the design was then tailored to the specifications using MATLAB. This involved inserting values for key parameters such as disk loading and mass, then checking the results.

The parameters to be input are listed below;

- Disk loading; this value is a measure of the aircrafts mass against the area swept from under its rotating lifting surfaces. Lower values of disk loading are preferable for the lower power requirements, but this value is a limiter for top speed and stability
- Velocity; the value was necessary to compare the characteristics of the aircraft in different speed regimes, however none of these speeds exceed 0.5 Mach (SL). This is due to the propulsion method not being capable at speeds over this value, as propellers are optimised for subsonic flight.
- Mass; this value was included as its was estimated already from existing data such as fuel consumption and mass of pilot.
- Power; this value was derived from the selected generator and engine combination, and would correspond to the maximum thrust able to be produced.

The parameters to be output are listed below;

- Rotor radius; this was a useful measure of the aircrafts compactness, and a smaller rotor radius is preferable for manoeuvrability concerns. However, to increase the endurance, a smaller radius is not ideal
- Top speed; measure of the top speed with max power delivered to flow
- Cruise speed; measure of speed at minimum drag
- Rate of climb; measure of the aircrafts ability to gain altitude, useful for comparing performance with other aircraft
- Endurance; a useful metric for finding the effectiveness of the craft at maintaining itself over an area

These parameters were used to test the motor response and the model simulation. The mass of the aircraft and the rotor radius were to be fed into the equations for modelling the motor and propeller.

## 2.3 3D modelling

During the process of parameter selection, the 3D model was also constructed. This process relied on a feedback between the parameters provided from MATLAB and the design choices kept in the model. Modelling the quadrotor required the least amount of preplanning, but the longest amount of time, due to the intricacies of using 3DExperience. The basic method involved was as follows.

- Construct the basic components needed, in very little resolution
- Form the basic assembly by placing the components in the correct places
- Refine those components which need refining (motor placements, wing aerofoil)
- Alter placement and characteristics as model develops
- Add materials and render

# 3 MAIN RESULTS

## 3.1 Configuration of aircraft design
 In this section, the 3D model is explained and several design choices are justified, along with listing of specifications and

### 3.1.1 Specifications

The specifications required from the tilt-rotor are below, and are drawn from a requirement based on existing examples in the field, although some are new. They are listed in order of importance.

- Lift production redundancy
- VTOL capability at low speed
- To be autonomously stable under the take-off condition (to augment the pilots control)
- To have a useful payload of 150kg or more
- To have a cruise speed of 175 km/h or more
- To have a top speed of 250 km/h or more
- Be able to cruise at less than half maximum thrust

### 3.1.2 General structure
The general structure of the aircraft, Fig. 1, consists of a frame (dark red), to which is mounted the components. One each edge is a wing section, connected to a duct (beige), which contains two motors(purple) and propellers(pink). In the fore section of the frame is the avionics(dark red), radar(grey) and pilot. In the aft section is the engine(yellow/red), fuel tank(orange), cargo bay(green), vertical tail (pink) and battery storage(grey). To cover the frame and components, a fuselage is mounted, with cut sections to allow the protrusion of wings and vertical tail.

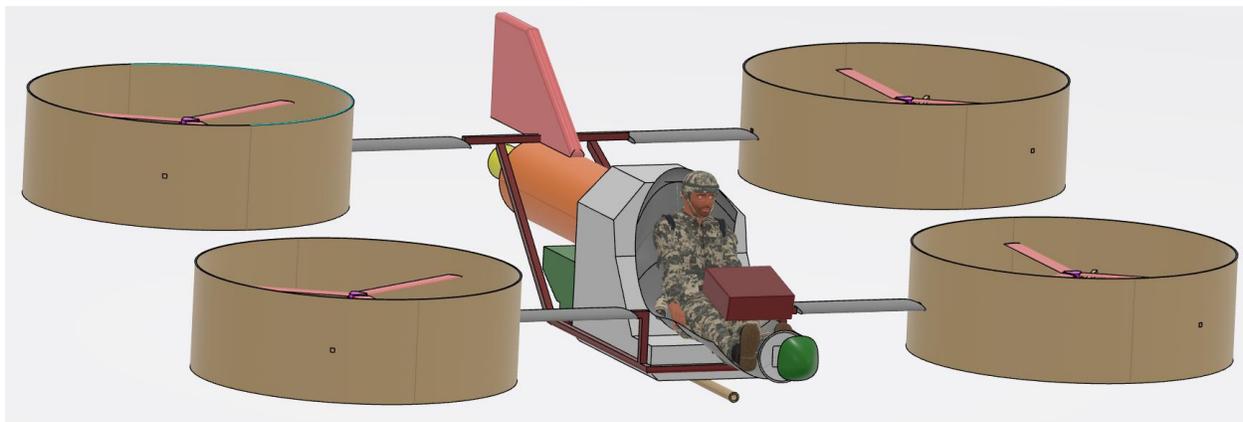

Figure 1 Aircraft without fuselage covering

Below is explained all the individual design choices used on the aircraft, including; Hybrid electric propulsion, Ducted fans, Tilting rotor shaft and wing sections, Duel propellers per rotor and Material selection.

### 3.1.3 Duel propellers per rotor

As mentioned in the introduction. The primary purpose of this new design is to add a failure safe design to the current VTOL aircraft market. During the literature review, many papers were discovered which proved the ability of a quadrotor to maintain altitude even with the loss of one rotor. This was achieved by inducting a rotating attitude about the z axis, sacrificing yaw controllability to maintain altitude, pitch and roll. While this is an effective solution for unmanned aircraft, when we considered the pilot experience during this failure state, it was assumed that too much disorientation would occur and inability to flight coherently will follow. As a result other alternatives were selected.

The presence of a wing allows for controlled flight during prop failure, as in this mode the aircraft operates as a convnetual fixed wing aircraft. However, the wing does not prevent critical failure during the VTOL condition. For this aim the solution to use a duel motor system per rotor was selected. While this does have the disadvantage of reducing the efficiency of flight, as disk loading is increase, it does increase top speed as frontal area is reduced, and the flow velocity difference between inlet and outlet is more significant.

### 3.1.4 Ducted fans

Ducted fans, shown in Fig. 2 were chosen for a wide variety of reasons:

1) Ducted fans increase propeller efficiency by removing tip losses

When the blade of a propeller generates thrust, a tip vortex is generated where the high pressure on lower edge combines with the low pressure on the top edge. This effect is called tip losses due to the reduction in useful thrust for the same power. The duct can reduce or remove this loss, depending on the clearance between blade tip and duct inner surface. As the blade tip is in close proximity to the duct, the tip vortex does not have room to form, and so the pressure difference between top and bottom surfaces is increased for the same power input, leading to increased efficiency. This leads to a higher top speed and increased range as a result of more thrust for the same power.

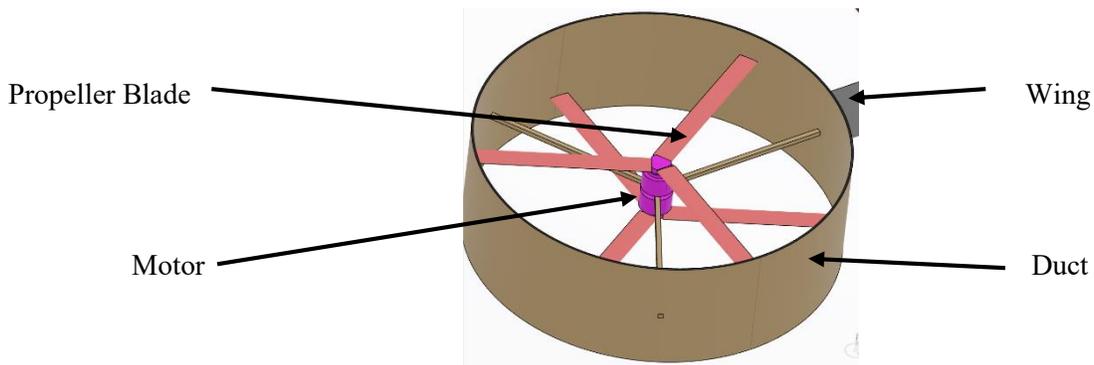

Figure 2 Ducted propellers of the aircraft

However, when the Angle of attack of the duct is too great the duct will stall and thrust falls quickly. This effect is mitigate by the duct being kept parallel with the flow direction by the tilting action. Due to the lift generated by the craft during cruise, the ducts can be more inline with the airflow. Without this lift, the craft would rapidly lose efficiency at speed.

2) Ducted fans protect motors and pilot from shrapnel and debris

The high profile of the duct allows for coverage of the motors, which provide the power to the flow. Assuming the ducts are made of a sturdy enough material, the coverage can help to protect the motors from stray paperiles and debris. The frontal ducts can also act as lateral protection for the pilot, as their placement covers the majority of the pilots body from side angles.

3) Ducted fans have a higher thrust for the same diameter as propeller

Due to the requirements that ducted fans have an asymmetrical number of propellers (nominally three), the thrust generated from the same area increases. A secondary benefit is the reduced frontal area of the duct when compared to a similarly designed propeller. This leads to a lower drag coefficient and as a result higher top speed. In combination with this, the duct thrust is increased by the efficiently boost of ducted tip losses, or lack thereof. The extra power added to the flow takes the form of extra thrust.

4) Increase ground safety

The aircraft benefits from the increased safety provided to ground crew, who may otherwise be in contact with high RPM blades. The duct provides a boundary between the crew and these blades.

The components of this assembly are:

- The motors (purple)
- The blades (pink)
- The duct (beige)

The motors are placed back to back to reduce the amount of supports needed to keep them connected to the duct. The blades are offset from the duct by 10mm to avoid collision at high blade speeds while also reducing the wing tip vortices. The duct is connected to the frame running through the wing sections, and will contain an actuator for tilting in future designs.

### 3.1.5 Tilting rotor shaft and wing sections

The tilting rotors were chosen to increase top speed by reducing drag. When compared to a conventional quadrotor, the tilt rotor is able to change the direction of the ducts and propellers without changing angle of attack of the fuselage. This allows the craft to convert the rotors from lift generators (under VTOL condition) to thrust generators (under forward flight condition).

A conventional quadrotor is reliant on a tilting of the entire fuselage to change the direction of the thrust vector. This has the small advantage of creating a lighter airframe, as no motors are needed to tilt the ducts. However this has the large disadvantage of papering a larger fuselage area into the direction of travel, where it will contribute significantly more to drag. Another disadvantage is the inability to maintain a constant angle of attack. This second disadvantage disallows a conventional quadrotor from using a wing of any kind. As a result the tilt-rotor was selected.

The addition of wing sections allows two significant benefits for this aircraft, in efficiency and safety. While the rotors have two propellers incorporated, the possibility of an engine failure is also a concern. With this in mind, the small wings can be used to keep the aircraft in flight despite total failure of either the rotors or the engine. In the case of total loss of a rotor, the aircraft can convert to conventional flight mode and thrust at half power until landing, increasing thrust on the damaged side and reducing on the undamaged side, while compensating with the rudder input. In the event that the engine shuts down, the aircraft can glide in the forward flight mode (poorly due to the low L/D ratio of 5.66), then during landing switch mode and use battery power to finalise decent.

### 3.1.6 Material selection

The aspect of the aircraft most interlinked with performance is mass. As with most materials used in aerospace, the materials used for this aircraft should be as light as possible. This will lead to the usage of the materials with the highest specific strength able to withstand the temperature to which it will be exposed. The modern materials which fill this category are composites. For this design to be successful, it should utilise as many modern composite materials that can fill the required role.

The main parts where materials can be selected are;

- The airframe: for the airframe, aluminium alloy will possibly be the best selection due to its low volume and high specific strength. Although carbon composites could be used, the requirement for a high survivability material led to aluminium alloys being considered more practical. As this is a low temperature application, the aluminium alloy will be safe from the effects rapid coarsening of their strengthening precipitates.
- The fuselage: This section is required to only cover the components of the aircraft. As a result, it should be made from Carbon fibre reinforced polymer composite, for its high specific strength.
- The propeller blades: Composite blades should be used due to its high specific strength and proven usage in engines such as the CFM International LEAP used on the Airbus A321neo. These blades will be lighter than the competing materials (aluminium alloys, steel, wood), while also having the benefit of utilising the fibres direction being inline with the force applied. This allows for even higher specific strength than the other uses of the material, which require fibre travelling in all directions.
- The duct should be constructed from Carbon fibre reinforced polymer also, but with added layers of protection for the blades. These would take the form of aramid fibre panels placed on critical spots on the ducts. This is due to the aramid fibres ability to absorb impact of blade fragmentation better than carbon fibres.
- The pilot seat and side panels should also be made from aramid fibre composites, to help protect from shrapnel impacts. Ultra hight molecular weight polyethene could also be a viable material for this purpose but could increase weight to such a degree as to be impossible to implement.

### 3.1.7 Hybrid electric propulsion
Hybrid electric propulsion was chosen for the tiltrotor.

1) Response time of brushless motors and failure protection

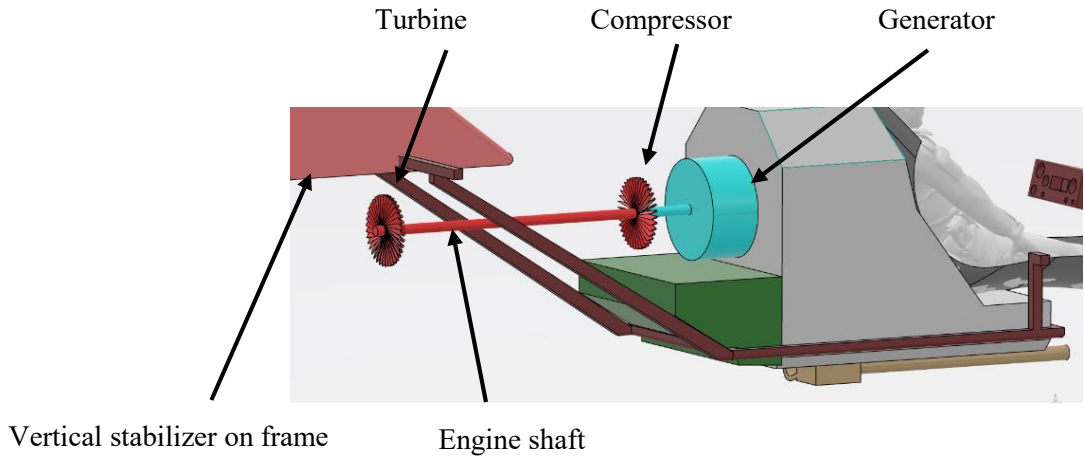

Figure 3 Hybrid propulsion (RED Engine shaft) (BLUE generator)

When selecting the primary propulsion method, most consideration was given to the take-off condition, as this was the most delicate design section. In order to be viable, the craft needed to have a response time fast enough for the pilot to deal with disturbances and collision avoidance. The options available were brushless motors or power shafts from the engine. Having a direct power shaft from the engine would require that all propellers rotate at the same RPM. As this could not generate a difference in thrust directly, a collective or variable pitch propeller would have to be fitted to each of the four rotors.

While this may have been an adequate solution, compensating for the loss of one of these rotors would have proved impossible, as this would require eight shafts and pitch variations. This increased complexity would offset any safety gained from using a second set of blades per rotor. As a result, electric motors were selected to drive the propellers. To drive these motors, a generator would be needed, the power for this generator would be provided by the engine shaft. This is shown in Fig. 3.

2) Efficiency can be kept by maintaining engine RPM

Most engines require a constant RPM for the most thermal efficiency. When using a hybrid system, the engine an maintain a constant RPM and allow any sudden changes in power requirement to be taken up by the batteries. The engine can then gradually increase power output, resulting in reduced fuel usage and increased endurance.

3) More-electric principles save weight and increase reliability and longevity

Using a hybrid powerplant allows for the utilisation of a wide variety of more-electric aircraft principles. These not only help to increase the weight savings and reliability, but also the lifespan and sustainability of the aircraft.

Using electric servos and actuators to control all movements of the aircraft dynamics sections can save a significant amount of weight. The hydraulics that would nominally power these movements contains a large amount of oil, which not only requires heavy high-pressure piping, but also the fluid itself is heavy. By replacing these with electric components, much weight can be saved. Increasing the reliability of the aircraft also benefits from the electric power arrangement. For the weight of one hydraulic line, several independent cables can run to the subsystem, allowing for damage mitigation, and protection from failure.

Using electrical power for most systems helps to future proof the design. As more effective power generation solutions are innovated, the craft can be retrofitted with this new system seamlessly, as all the systems that would nominally require a conventional engine to operate have been replaced in the design stage.

4) Batteries low specific energy

In recent years battery technology has been constantly innovated and improved upon, leading to batteries with greater characteristics in specific energy and power. However, despite these advantages, batteries still remain uncompetitive in high energy aviation applications. JP-1, will typically display specific energy values of 43MJ/kg while the most promising batteries will only manage around 0.5MJ/kg. This stark difference in specific energy, with hydrocarbon fuels being greater than eighty times more effective, led to the conclusion that for such a highly powered aircraft, the only solution was hydrocarbons. Using the same mass of batteries as fuel and powerplant combined would only provide a flight time of 22 minutes at cruise [14]. Considering the increased efficiency during flight as the hybrid aircraft loses mass via combustion of fuel, batteries are not viable for a high speed craft of this type

### 3.1.8 Specifications

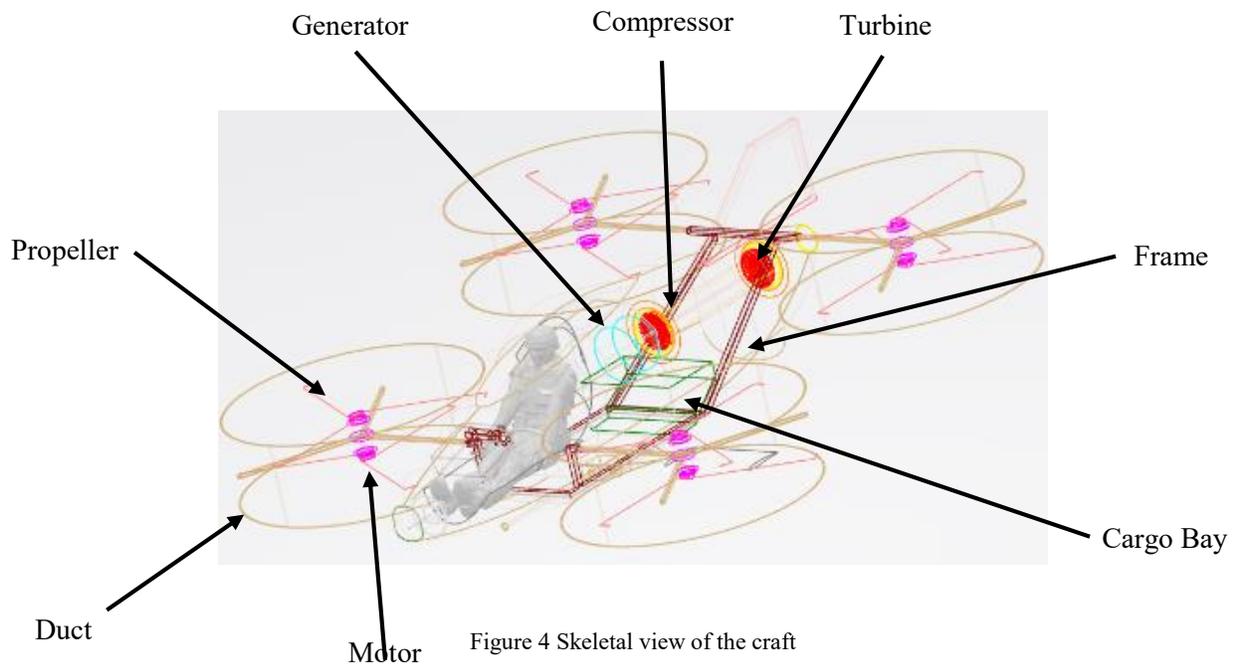

Figure 4 Skeletal view of the craft

Figure 4 shows a skeletal view of the aircraft, displaying the internal components in the correct place, while being able to see the main structure as well. Table 1 shows the calculated specifications for the design, of which some are used for the simulation.

**Table 1**
**Aircraft specifications**

| Value name | Value | Notes |
|---|---|---|
| Crew | 1 | Could carry sling with casualty without arms |
| Capacity | 101 (kg) | Arms, cargo and pilot excess mass |
| Length | 6.1 m | |
| Width | 6.7 m | |
| Height | 2.3 m | |
| Wing Area | 3 m$^2$ | Not including body lift area |
| Empty weight | 265.6 kg | No fuel, pilot, arms or cargo |
| Operating weight | 577 kg | Max safe weight, where can climb despite motor failure |
| Max vertical take-off weight (not failure safe) | 778kg | Full power delivered from engine |
| Fuel Capacity | 119.5 kg | JP-1 |
| Powerplant | 340kW | Hybrid power train |
| Rotor diameter | 1.5m | |
| Rotor area | 1.71 m$^2$ | 4 of these leads to total area of 6.84 m$^2$ |
| Wingspan | 3 m | Divided into 4 sections |
| Aspect ratio | 3 | |
| V stall | 43 m/s | Cl=1.25 |
| V Cruise | 70 m/s | Cl=0.60 |
| V High Cruise | 93 m/s | Cl=0.29 |
| V Max | 120 m/s | Cl=0.17 |
| Max range | 957.00 km | Travelling at L/D max |
| Sealing | 3,080 m | Hypoxia risks |
| Maximum glide ratio | 5.66:1 | |
| Wing loading | 192.3 kg/ m$^2$ | |
| Rate of Climb | 10.72 m/s | |
| Rate of Climb Max | 31.82 m/s | |
| Power deliver to flow/mass | 220 W/kg | |
| Armaments | 1 hardpoint for mounting items with total mass less than 63kg. Nominally 20mm Autocannon. | |

### 3.1.9 Iterations of 3D model

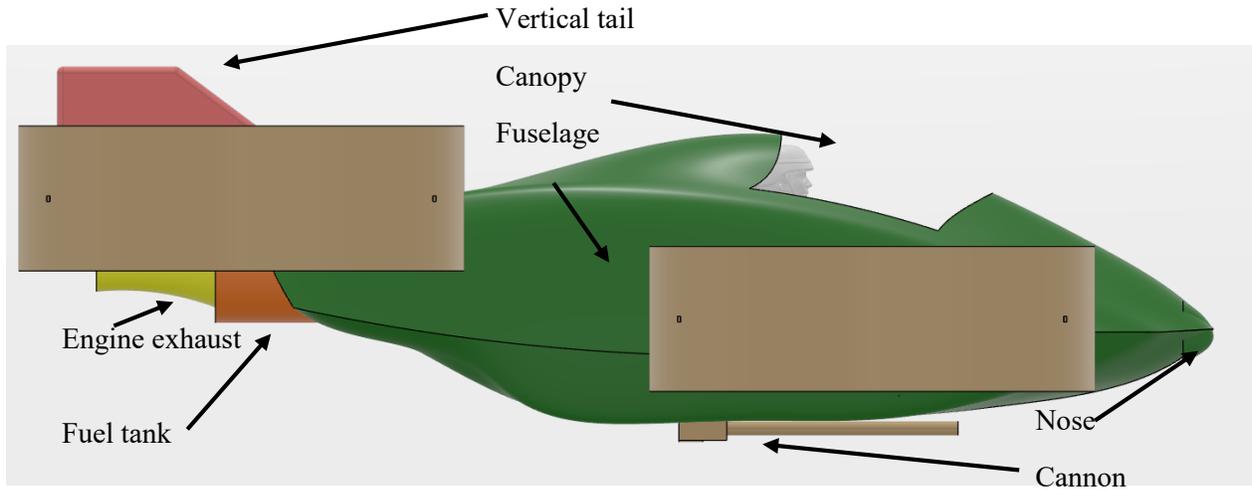

Figure 5 Side view of first iteration

Figures 5 and 6 show the first iteration of the aircraft.

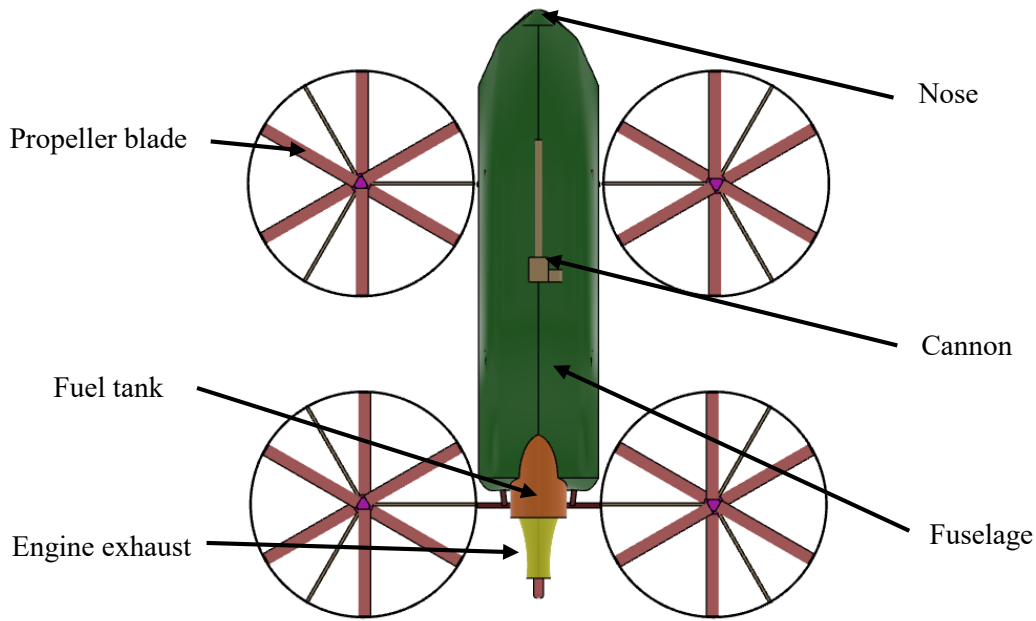

Figure 6 Lower view of the first iteration

The initial placement of the rotors was selected to reduce the width as much as possible. This would have the advantage of reducing moment of inertial about the aircraft longitudinal axis, making the craft more manoeuvrable. Also, this would allow the craft to fit into a smaller hanger. The initial fuselage was created by several splines running the length of the aircraft, with a focus on reducing the frontal area as much as possible. This design had the benefit of a reduced frontal area, however the construction of such a fuselage would create a variety of manufacturing issues due to this design method. Therefore, in the second iteration, the fuselage was changed.

This first iteration also left the latter sections of the fuel tank and engine exposed. This was done to save weight from the fuselage component. However, this led to a significant reduction in aerodynamic shape. Figure 6 shows this well, as we see the square section of fuselage where the engine is exposed. The placement of internal components was such that the centre of mass was close to the centre of the cross formed between the four rotors.

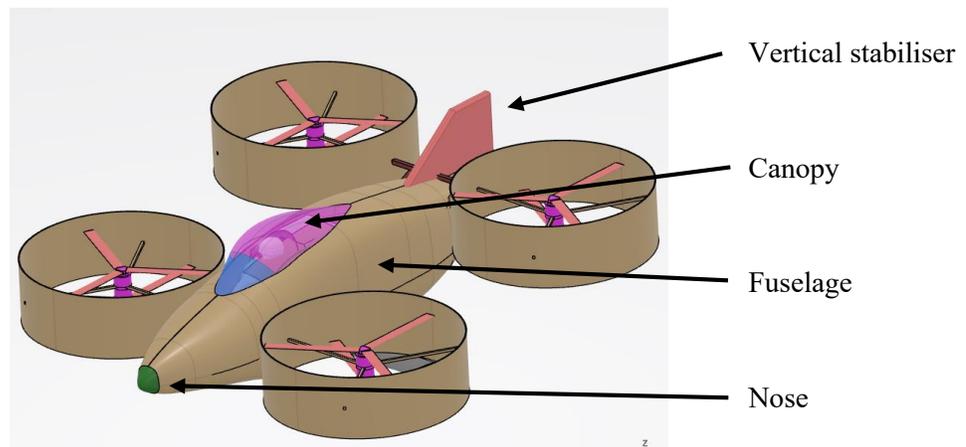

Figure 7 Total view of second iteration

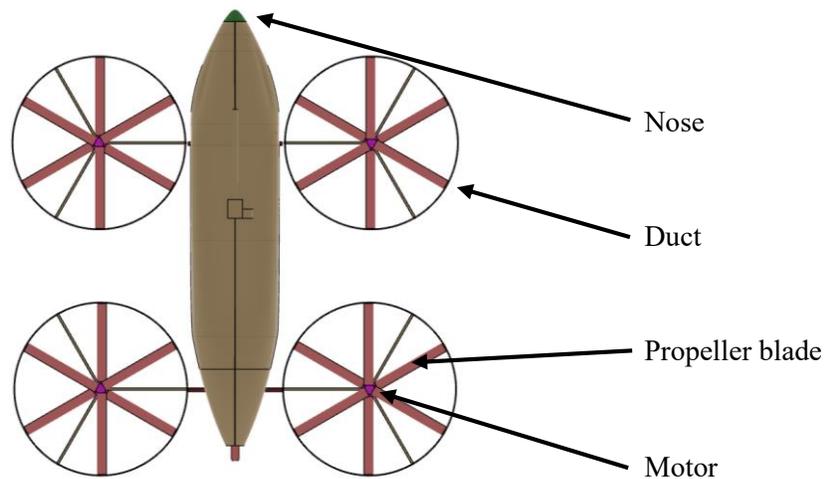

Figure 8 Bottom view of second iteration

While creating the second iteration, Figs 7 and 8, several changes were implemented. The first and most obvious is the change in the fuselage shape and construction method. The shape change was informed by aerodynamic reasoning, with the addition if a streamlined section to the rear, covering the fuel tank and engine. Only a small hole is left for the exhaust. Also, the shape has been elongated to smooth the transition between increasing area, thereby reducing the drag coefficient.

The construction method change was does in two parts. The first was the use of conics and lines as opposed to the splines used in the first iteration. This new method was useful, as it allows to easily change the parameters of the curves which make up the aircraft structure, of which the fuselage surfaces are constructed from. As a result, easy tailoring of the of fuselage shape was achieved. The second construction method used was to make the canopy and fuselage separate entities. This allowed for the smallest amount of fuselage material, while allowing the pilot a wide range of vision. As the canopy is made from a separate part, manufacture will be eased during production.

The other minor changes include shifting of the pilot seat and fuel tank to move the COG further towards the centre of lift under the vertical take-off condition

The rotor placement and construction remained the same in the second iteration, as this was one of the first assembly to be properly made to calculated specifications.

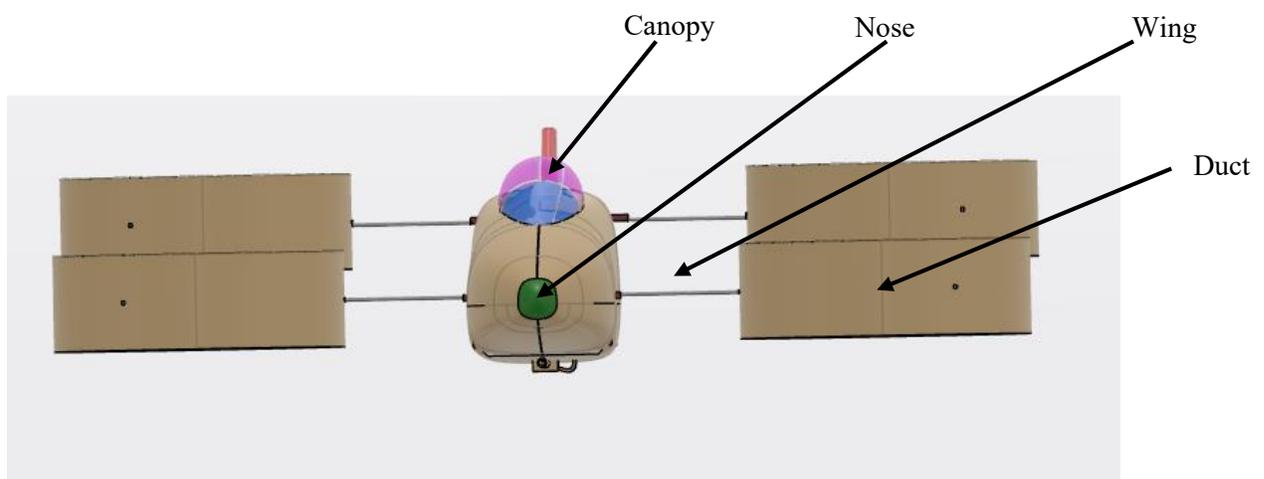

Figure 9 Offset of frontal view of third iteration

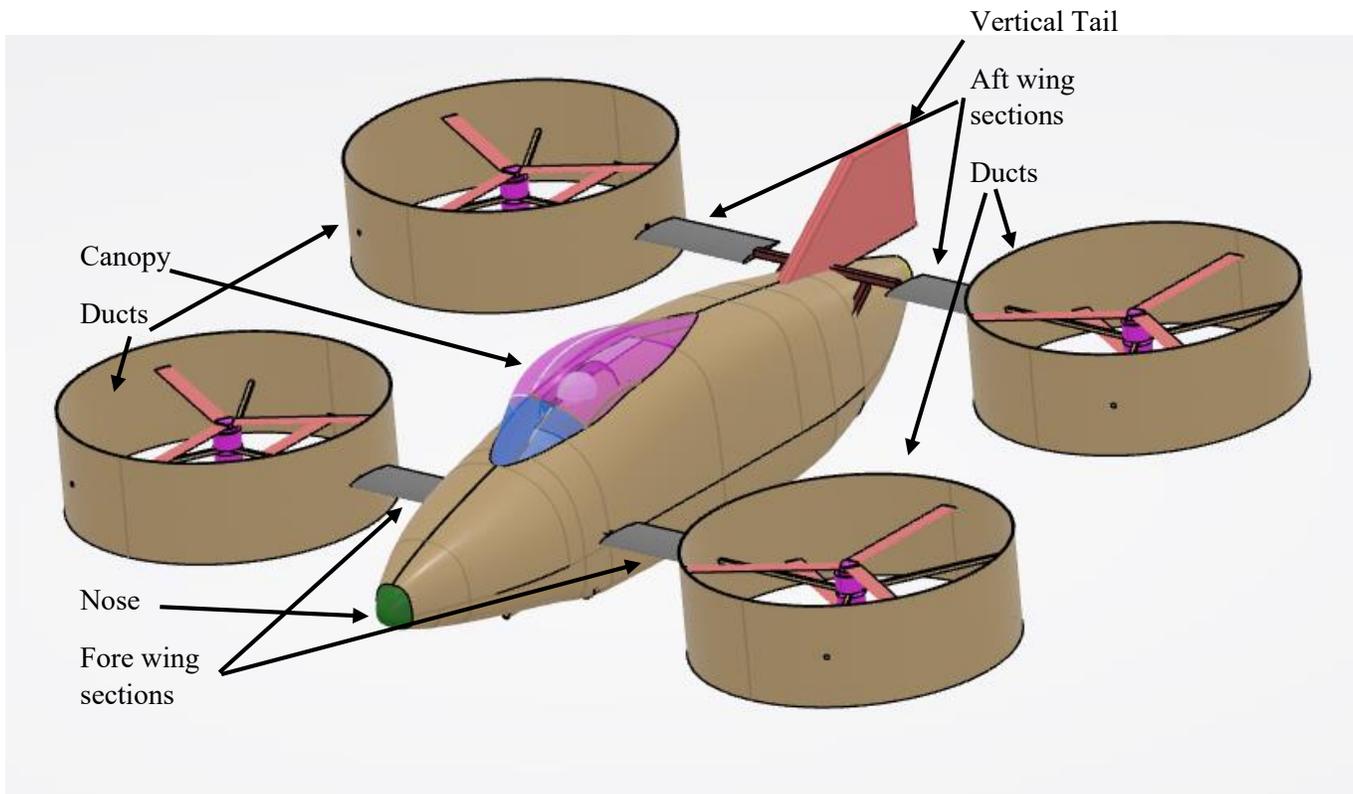

Figure 10 Total view of the third iteration

The third and final iteration for this paper is shown by Figs 9, 10 and 11. When compared to the last iteration, the only significant change is the addition of wing sections. Due to the thin aerofoil selection (as only considered for cruise) no components can be stored in these short wing sections. As a result, the wing sections will be constructed from one of the Carbon Fibre Reinforced Polymer materials. The central spar running through will transmit the load to the fuselage and be made of an aluminium alloy.

These wing sections do change the shape of the craft somewhat, leading to a shape wider than initially considered. However, all the negatives produced by the additional width are offset by the large increase in to speed, as the ducts can be fully tilted into the flow direction, proving all power to the flow and minimising drag.

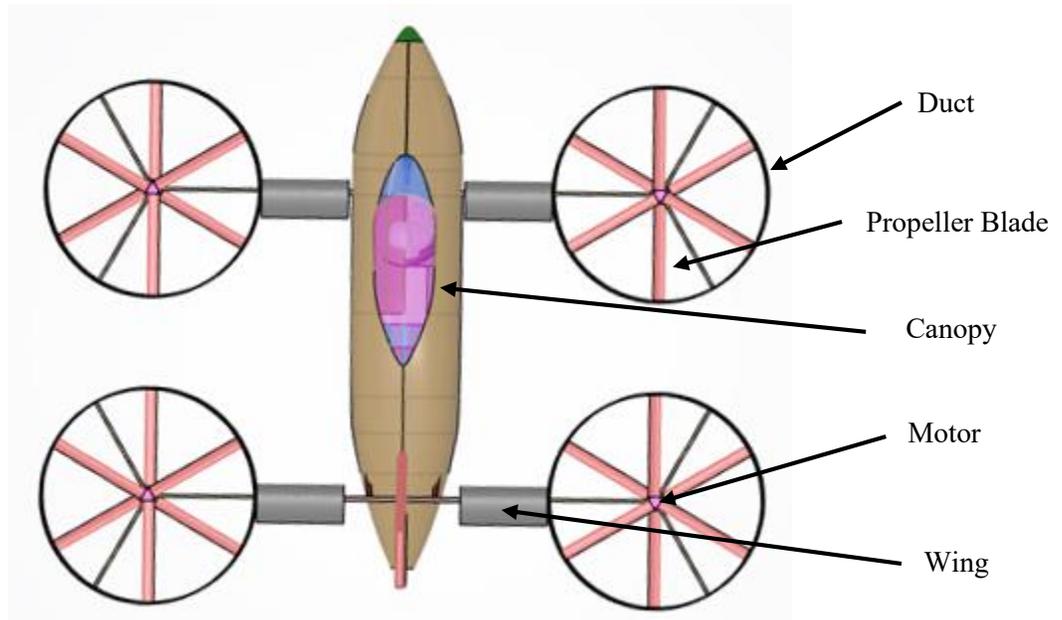

Figure 11 Top View of Third iteration

## 3.2 Aircraft model

### 3.2.1 The detailed mathematical model

This mathematical model was derived from the basic dynamics of an aircraft experiencing a force shown below.

$$ma = F = m\ddot{x} = f(t) - \lambda_{up}\dot{x} \qquad \ldots(5)$$

This then converted into the s domain is.

$$F(s) - s \cdot \lambda_{up} X(s) = m \cdot s^2 X(s) \qquad \ldots(6)$$

This can be rearranged into the transfer function

$$\frac{X(s)}{F(s)} = \frac{1/m}{s^2 + s \cdot \frac{\lambda_{up}}{m}} \qquad \ldots(7)$$

With input of Force and output of displacement.

we can combine with the motor model found in Section 1.5

$$\frac{F(s)}{V(s)} = \frac{K_m}{L_f \cdot s + R_f} \qquad \ldots(8)$$

Combining (7) and (8) lead to

$$\frac{K_m}{L_f \cdot s + R_f} \times \frac{1/m}{s^2 + s \cdot \frac{\lambda_{up}}{m}} = \frac{K_m/m}{(s^2 + s \cdot \frac{\lambda_{up}}{m}) \times (L_f \cdot s + R_f)} \qquad \ldots(9)$$

$$\frac{K_m/m}{(s^2 + s \cdot \frac{\lambda_{up}}{m}) \times (L_f \cdot s + R_f)} = \frac{K_m/m}{L_f s^3 + (\frac{\lambda_{up}}{m} + R_f)s^2 + s \cdot \frac{\lambda_{up}}{m} R_f} = \frac{X(s)}{V(s)} \qquad \ldots(10)$$

This model, equation (10) takes an input of voltage and outputs displacement.

### 3.2.2 Controller formula
The control formula was derived from the closed loop transfer function;

$$CLTF = \frac{G_s}{1 + G_s} \qquad \text{...(11)}$$

Were

$$G_s = G_c G_0 \qquad \text{...(12)}$$

And

$$G_c = k_d\, s + k_p + k_i \cdot \frac{1}{s} \qquad \text{...(13)}$$

$$G_0 = \frac{K_m/m}{L_f s^3 + (\frac{\lambda_{up}}{m} + R_f)s^2 + s \cdot \frac{\lambda_{up}}{m} R_f} \qquad \text{...(14)}$$

This rearranged leads to the characteristic equation of.

$$\frac{1}{L_f}[L_f s^3 + (\frac{\lambda_{up}}{m} + R_f)s^2 + s \cdot \frac{\lambda_{up}}{m} R_f + \frac{K_m}{m} \times (k_d\, s + k_p + k_i \cdot \frac{1}{s})] \qquad \text{...(15)}$$

Or

$$\frac{1}{L_f}[L_f s^4 + (\frac{\lambda_{up}}{m} + R_f)s^3 + s^2 \cdot \frac{\lambda_{up}}{m} R_f + \frac{K_m}{m} \times (k_d\, s^2 + k_p\, s + k_i)] \qquad \text{...(16)}$$

Leaving

$$\frac{1}{L_f}[(L_f) \cdot s^4 + (\frac{\lambda_{up}}{m} + R_f) \cdot s^3 + (\frac{\lambda_{up}}{m} R_f + \frac{K_m}{m} k_d) \cdot s^2 + (\frac{K_m}{m} k_p) \cdot s + \frac{K_m}{m} k_i] \qquad \text{...(17)}$$

This leads to the 4$^{th}$ order polynomial function

$$s^4 + kb\,s^3 + kd\,s^2 + kps + ki = (s - s1)(s - s2)(s - s3)\,(s - s4) \quad \ldots(18)$$

Were we select the poles (s1-s4) based on response. These regulate the values of kd, kp and ki, where kb is based on the unchangeable parameters of the motor and model.

### 3.2.3 Basic model block

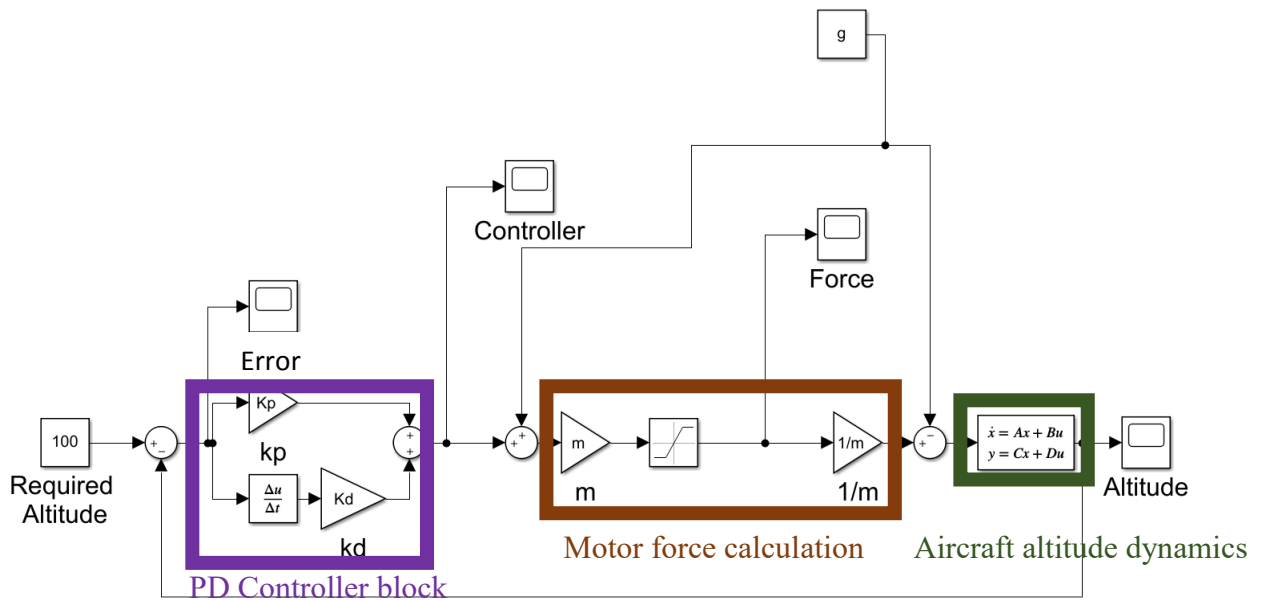

Figure 12 Basic control system to analyse control force

To test the basic requirements of control force and the effect of motor saturation on aircraft dynamics, this control system was constructed. Figure 12 displays this. In the purple box, the PD controller is displayed, which takes the error value and converts it to an acceleration value. The brown box represents the motor force calculation, with the added saturation which represents the motors maximum take-off force. Finally the green box contains the aircraft altitude dynamics, converting the acceleration input into a value of vertical displacement, to represent the present altitude. The values in Table 2 are the parameters for the system.

Table 2
Basic control system values and effects

| Variable name | Nominal Value | Effect |
|---|---|---|
| m | 577 | Represents the mass of the aircraft |
| Kp | 0.65 | The proportional gain of to the current error |
| Kd | 5 | The gain to the derivative of error |

### 3.2.4 The control system including motor

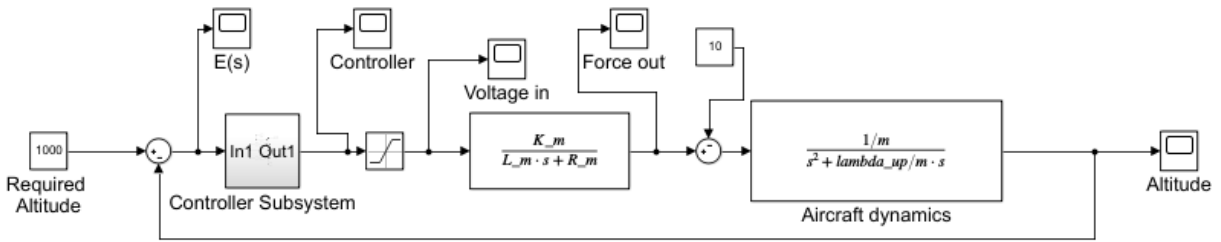

Figure 13 Altitude Dynamics Control system

figure 13 displays the control system used to measure the dynamics while including the motor model. The model can be broken down into 3 sections, shown below.

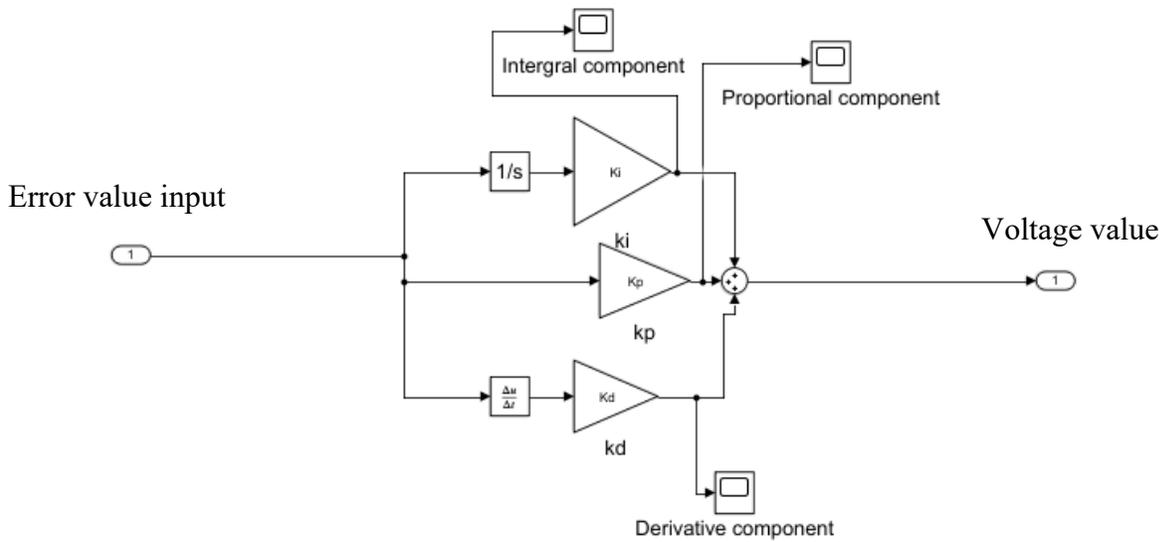

Error value input

Voltage value

Figure 14 The controller section

**Table 3**
**PID controller parameter values and effects**

| Variable name | Nominal Value | Effect |
|---|---|---|
| Kp | 4.142 | The proportional gain of to the current error |
| Kd | 30.48 | The gain to the derivative of error |
| Ki | 0.004 | The gain to the integral of error |

The first section, the controller in Fig. 14, contains the PID controller. This controller contains three different methods for making the error tend to zero. The first of these is the proportional term. This term takes the error value, or the difference between the value required and the value currently, multiplies it by a gain, then delivers the result. The gain for this section is Kp. The next value is the derivative gain, which takes the derivative of the error and multiplies it by a value of Kd. This acts to avoid oscillations in the other signals. Lastly is the integral value, which integrates the error remaining by the value of Ki. These three signals combine to provide a value for the control voltage, which is then fed into the motor model. Table 3 contains all the parameters currently used in the system

**Table 4**
**Motor model values and effects**

| Variable name | Nominal Value | Effect |
| --- | --- | --- |
| m | 577 | Represents the mass of the aircraft |
| K_m | 10 | Gain relating Voltage to Force |
| L_m | 0.110 | The inductance of the selected motor |
| R_m | 0.140 | The resistance of the selected motor |
| Lambda_up | 9 | The air resistance of the aircraft in the |

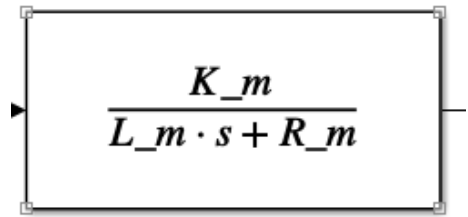

Figure 15 The motor model (Simplified)

This model was chosen based on research shown by [15]. However the model seen in Fig. 15, was to be simplified in order to allow for ease of control. The voltage output from the controller is fed through this block then converted into the force delivered by the motors. The usage of a gain to represent the transformation of the nominal torque output from the motor into voltage is where the simplification occurred. The values of the inductance, L_m and resistance, R_m were taken from the motor specification sheet to provide realistic responses to the input values. Table 4 contains the parameters used in this system.

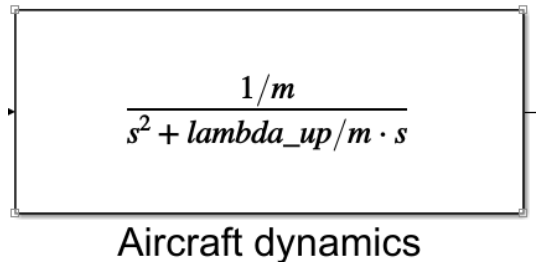

Figure 16 s domain aircraft dynamics and block internals

Figure 16 describes the aircraft altitude dynamics of the quadrotor. This block was necessary to convert the force provided from the motors into the movements of the aircraft in 3D space. Later this model will be explored in further detail in order to fully understand each components necessity in the system.

### 3.2.5 The control system including motor for the attitude dynamics

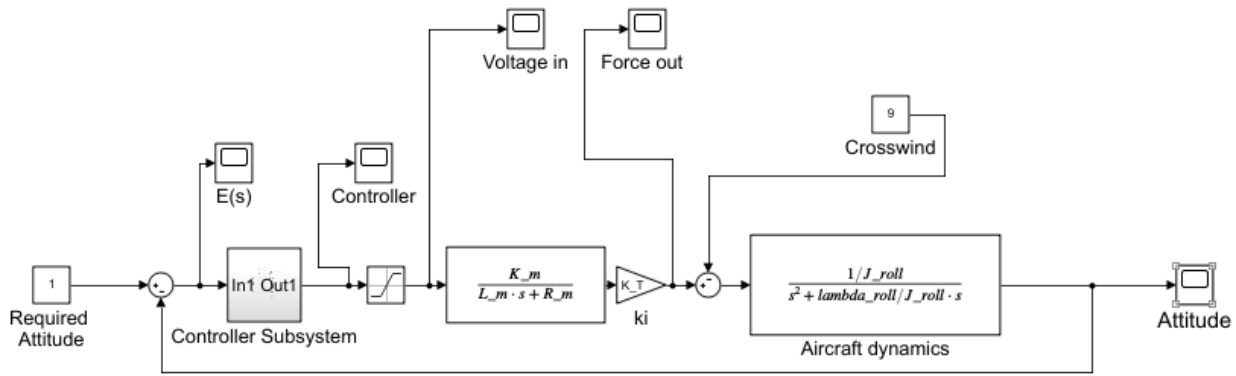

Figure 17 Roll dynamics Control system

Shown in Fig. 17 is the control system including the motor block for the attitude dynamics . This included a higher saturation as the motors were able to use more of the potential power, and also the addition of the constant gain K_T, which represents the change from force into torque due to the distance of the forces from the COG. On the left the controller takes the value of current roll angle error and outputs the voltage. This voltage, once passed through the saturation block, is fed into the motor dynamics block, where the torque value is output. This torque is input into the aircraft dynamics block, where the roll dynamics are represented. Finally the actual value of roll angle is output and compared with the required value to find the roll angle error. Table 5 contains the addition constant required for this system.

**Table 5**
**Torque constant value and effect**

| Variable name | Nominal Value | Effect |
|---|---|---|
| K_T | 1.2 | Converts the force from the motors into a torque about COG |

## 3.3 Aircraft performance.

This section is intended to display the capabilities expected from an aircraft of this type, such as top speed and endurance. These parameters are only the result of preliminary analysis and as a result are only estimates of the performance able to be delivered. The parameters for analysis are displayed in Table 6

**Table 6**
**Parameter table for aircraft performance**

| Symbol | Name (units) | Use |
|---|---|---|
| P | Power (W) | Determine load on engine and generator, can be used to calculate fuel requirements |
| m | Mass (kg) | The total mass of the aircraft under flight conditions (Max take-off for safe operations) |
| E | Endurance (hours) | To determine the length of time in the air |
| dl | Disk loading (kg/m^2) | The amount of mass the aircraft supports per unit area of lift generators |
| V | Velocity (m/s) | The airspeed of the craft |
| T | Thrust (N) | The force delivered by the propellers to the air to produce a pressure difference |
| D | Drag (N) | The force acting on the aircraft due to its speed through fluid |
| RoC | Rate of climb (m/s) | The speed which an aircraft can climb, the excess power delivered to gaining altitude |
| FF_rate | Fuel flow rate (kg/s) | The rate of fuel taken in by the engine |
| R | Range (km) | The range which the aircraft can fly in a straight line |

### 3.3.1 Disk loading selection

The disk loading of the aircraft was optimised for several reasons.

Disk loading is just a measure of the aircrafts mass to lift generator area, show below

$$DL = m/A \qquad \ldots(19)$$

However, this value has wider implications, as when disk loading increases, the velocity of the flow must also increase to provide the necessary momentum change to keep the aircraft in a stable hover. This velocity increase does not linearly relate to disk loading, shown in equation (20).

$$v = \sqrt{DL \cdot \frac{g}{2\rho}} \qquad \ldots(20)$$

And this relates to the power required via equation (21)

$$P = Tv = mg\sqrt{DL \cdot \frac{g}{2\rho}} \qquad \ldots(21)$$

A higher disc loading led to;

- Increased power usage
- Increased top speed
- Lower endurance
- Increased stability
- Increased response time

As a result, a balance is to be struck between these effects depending on the vehicle requirements. We need to maintain a realistic power draw for a vehicle of such a small size, and reducing the engine size helped to create a positive reduction in mass.

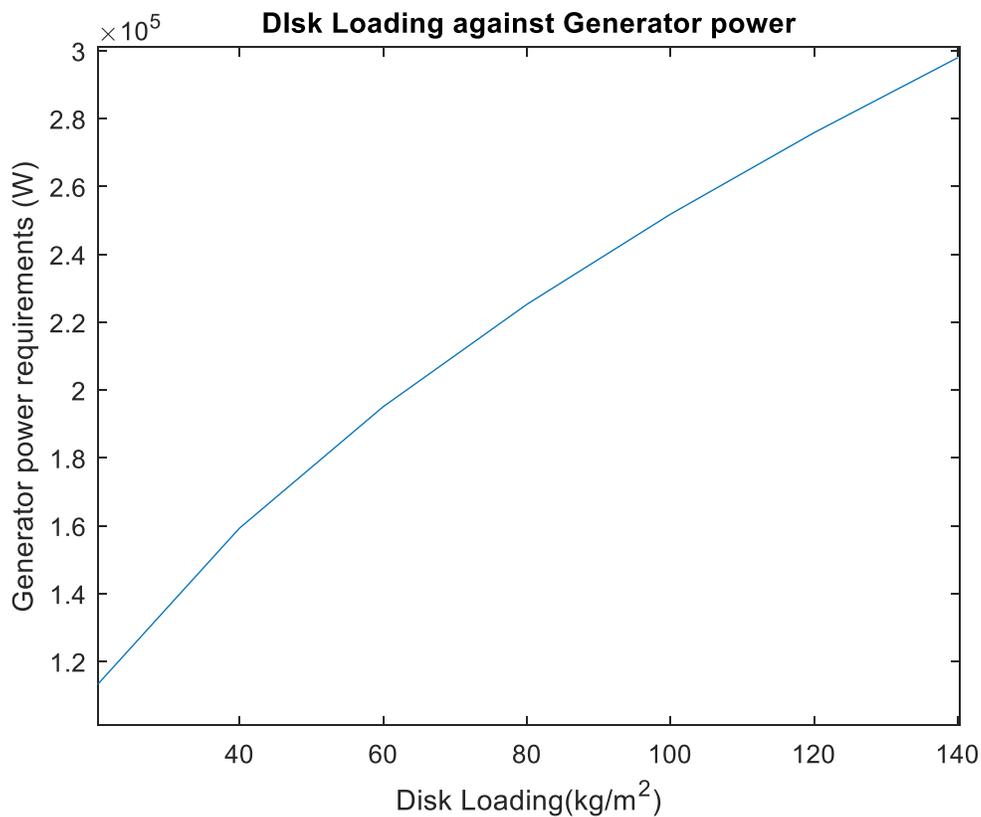

Figure 18 Disk loading verses the power required from the generator during accelerated hover

Figure 18 shows how the power required for the same load constraint (accelerating hover). It can be seen that as the disk loading increased, so does the power required from the generator. Notice that at a loading of 120kg/m², the generator must produce a power of 270kW, whereas at a DL of 80kg/m² this value drops to 220kW. This demand scales with the power required from each rotor, as is it can be reduced during the hovering sections, then the excess power can be used for increased manoeuvrability.

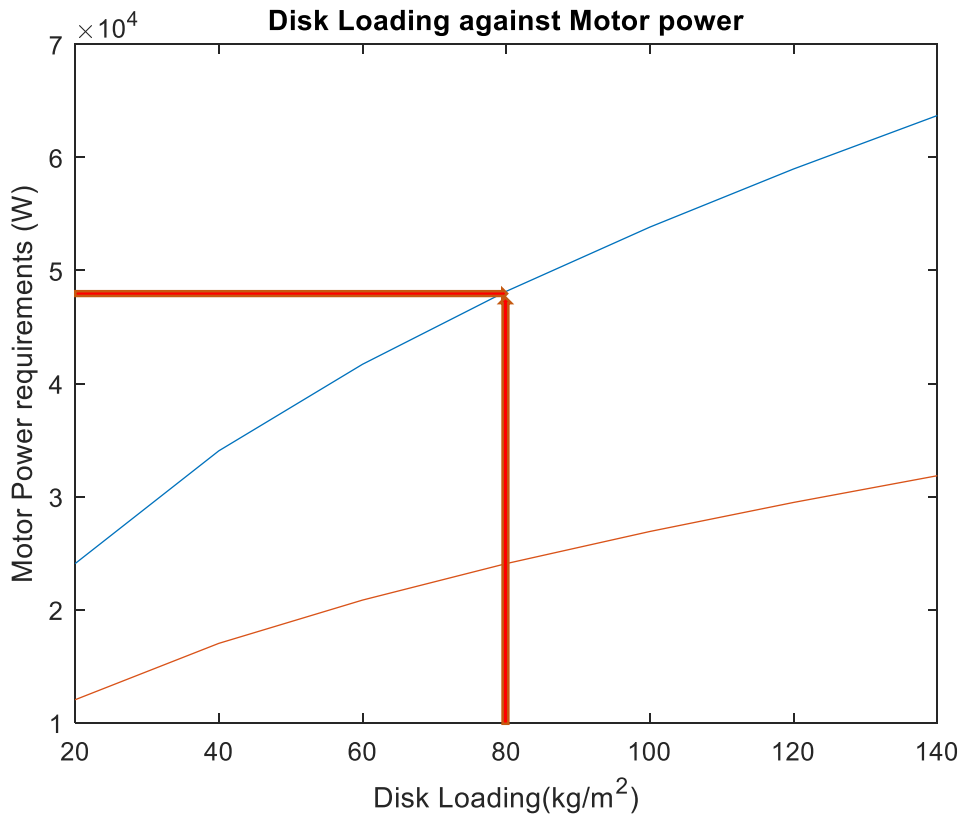

Figure 19  Motor power draw against disk loading for take-off (0.2g)

BLUE (MAX POWER) ORANGE (Standard)

Figure 19 shows the difference in power requirements under two conditions of the flight. The orange line represents the aircrafts propulsive motors power draw when operating under normal conditions, where both propellers in the ducts are operating. The blue line displays the effect when one of the motors or propellers has stopped working, and the other must provide all the power to the airflow. It can be seen the large difference between these power requirements, and this is where the limiting factor of the aircraft size is displayed.

Due to the higher power requirements for such a vehicle, the motors currently available must be able to produce this power, with a reasonable response time. As it stands, the motors selected for this task were the MP154120 48KW. These motors weight only 6kg and deliver 48kW of peak power, and are already proven in the Chinese aerospace industry. If the power required from the motor exceeds this value, then the aircraft will not be able to climb while using one motor in the vertical flight configuration. The red arrows on Fig. 19 show the maximum disk loading for which the aircraft can climb(0.2g acceleration included in power requirements) under the failed state. If this disk loading is exceeded, then the aircraft will not be able to continue mission.

It should be mentioned that this is a worst-case scenario, as during the flight, fuel will be burned, and the DL reduced. With this in mind, if the DL is kept at 80kg/m², the aircraft is safe to fail under all flight conditions, including full weight vertical take-off. As a result the full load disk loading was chosen to be no more than 80kg/m². This leads to a propeller radius of 0.75 meters.

### 3.3.2 Mass breakdown

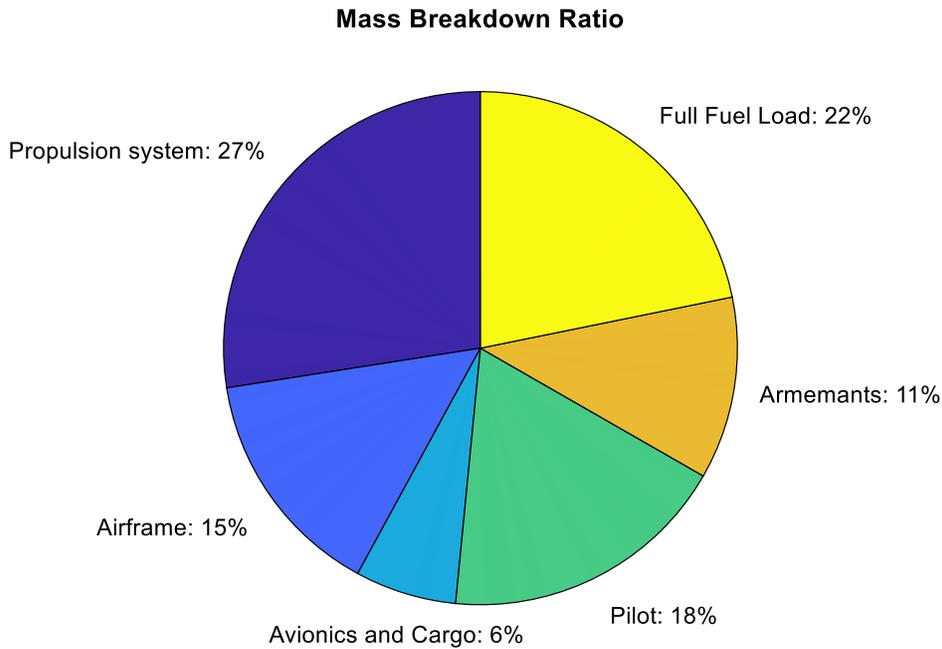

Figure 20 Aircraft mass breakdown

**Table 7**
**Aircraft mass estimates**

| Name | Mass (kg) (estimate) | Notes |
|---|---|---|
| Pilot | 100 | Weight of human with flight equipment and parachute |
| Fuel | 119 | Full fuel load |
| Arms | 63 | Defensive equipment |
| Propulsion | 150 | Includes motors generators engine and cooling |
| Airframe | 80 | Fuselage and structure |
| Misc | 35 | Avionics Radar cargo |
| Total | 577 | Total aircraft mass |

The total mass of the aircraft can be divided into several categories and are shown in Fig 20 and Table 7. The preliminary estimations for each of these categories are as follows. For the pilot, the average weight of a person was used, which came to a value of 81kg. we then considered the additional equipment needed by a pilot, such as flight suit, parachute, survival equipment, and other possible supplies. An estimated 19kg.

The fuel value was based on existing examples, by analysis the empty weights of existing helicopters and VTOL aircraft and finding a ratio between empty weight and fuel capacity. This led to an average percentage of 45%. As our estimated empty mass changed, so did the fuel load. When the final empty mass of 265kg was found, the fuel load was then 119kg.

The mass of arms was calculated by looking at the masses of existing equipment used by militaries. The example chosen was the 20mm MG151 (42kg) with 200 round of ammunition(20kg), leading to a total mass of 63kg. Again this is a generous estimate, as most likely armament used would be a lighter than this value.

The propulsion system was estimated through a continuous cycle of iteration, using existing technology as the basic estimate. After many iterations the motors were considered to be 6kg (48kg total), the generator to be (16kg) and the engine to be (62kg). this also includes the mass of the propellers, which came to 3kg each.

The airframe consisted of the estimate with the most varied mass, however, using existing helicopter masses as example, the mass of 80kg was settled on. This mass considers the frame, the ducts and the fuselage skin, along with the canopy.

Finally, an extra 35kg was given to provide mass for the electronic systems, such as the radar, avionics, radio and cargo. This estimate was based on the 0.5-3% weight of modern aircraft being made of electrical systems and avionics [17].

### 3.3.3 Flight velocity, Rate of climb and Endurance

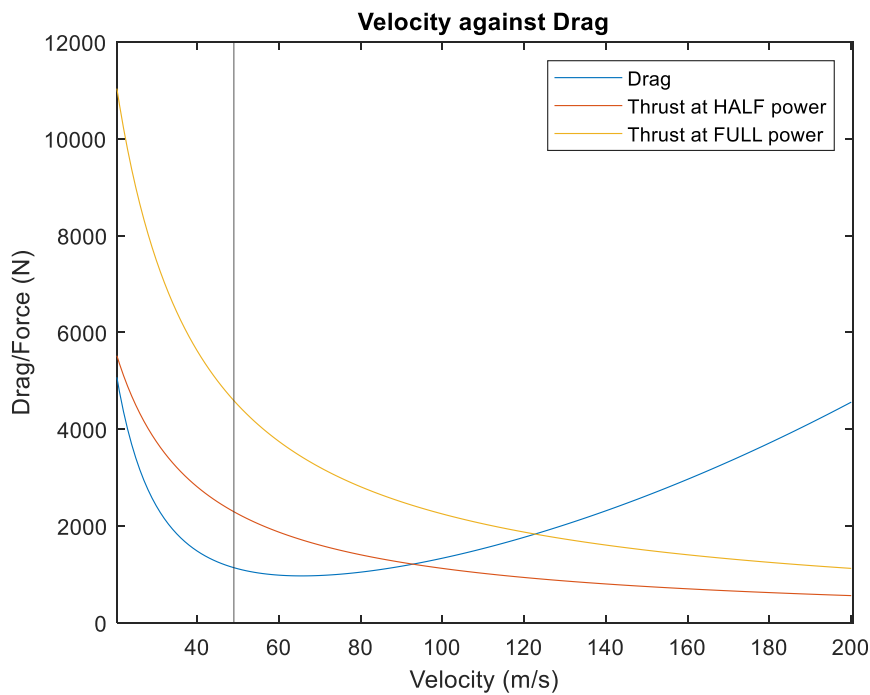

Figure 21 Graph showing the thrust and drag at different power settings

**Table 8**
**Aircraft flight speeds and related power, endurance and range**

| Mode name | Forward velocity (m/s) | Drag value/Force required (N) | Power required from generator (after inefficiencies) (kW) | Fuel usage rate (kg/s) | Endurance estimate for whole mode (hours) | Range (km) |
|---|---|---|---|---|---|---|
| Min power | 52 | 1077 | 89.28 | 0.0071 | 4.97 | 931.1 |
| Cruise | 70 | 970 | 108.30 | 0.0087 | 3.80 | 957.6 |
| High Cruise | 93 | 1210 | 179.47 | 0.0140 | 2.37 | 793.5 |
| Top speed | 120 | 1768 | 338.4 | 0.0207 | 1.26 | 542.8 |
| Hover | 0 | 5660 | 225.27 | 0.0175 | 1.90 | 200 |

Figure 21 and Table 8 show the aircraft in different flight conditions. Min power was the condition where the least power was used, which allowed for maximum endurance. This flight mode would be useful for search and rescue operations, to allow for increased time over the target area. Cruise is the position where drag is minimum, and as a result, range is highest. This mode is the most efficient speed to travel from location to location, if no other factors are considered. High cruise is where the motors are at half power and is useful for casualty extraction and other long distance but time critical missions. Top speed represents the maximum power output by the generator. To be used as emergency power were speed is critical. [18] These values are all well above the specifications set in section 3.1.1, as the cruise speed at half power is 305 km/h, and the top speed is 393 km/h.

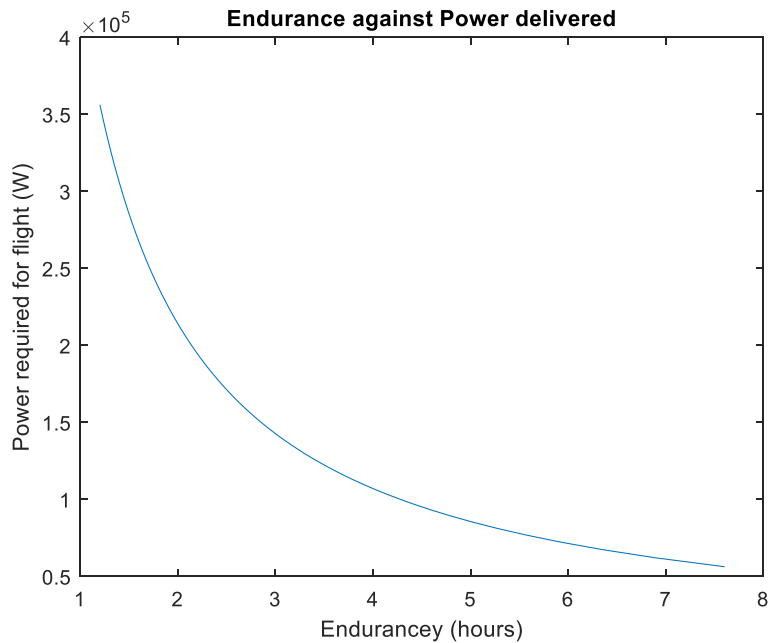

Figure 22 the endurance of the craft against the power drawn

As Fig. 22 shows, the endurance that can be achieved is directly linked to the power drawn by the propulsion system.

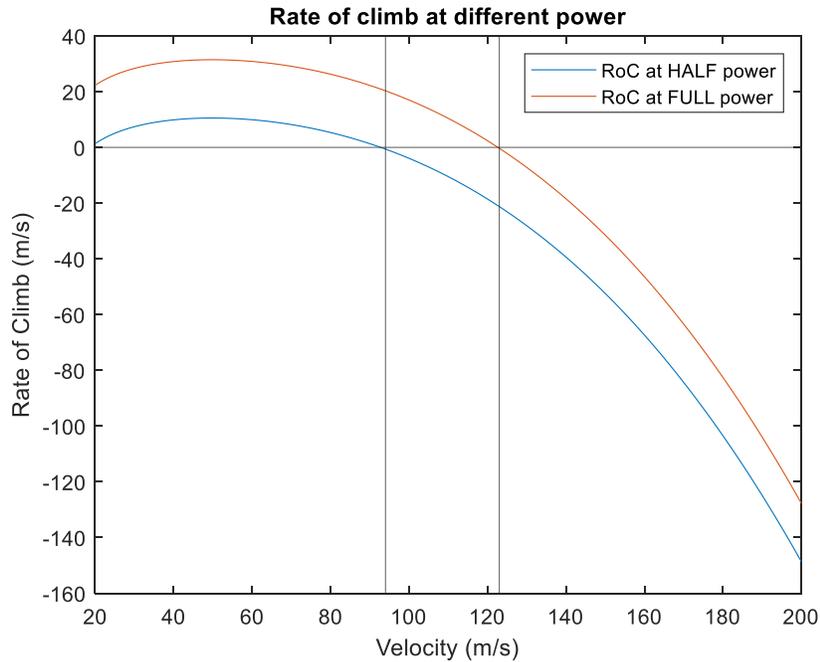

Figure 23 Rate of climb at different power settings

This graph, Fig. 23, shows how much excess thrust is produced at different speeds. The two vertical black lines show the maximum velocity at the respective speeds, or in this case points where Rate of Climb is zero. We can see that the aircraft can climb under all conditions below the top speeds. This is a good metric to determine horizontal manoeuvrability as the aircraft can maintain altitude regardless of current rotor configuration. If the craft were stalling, the ducts could be tilted to an angle to provide the extra lift needed to maintain level flight.

# 4 SIMULATION PART

## 4.1 Simulation Results

The simulations below were conducted on the MATLAB program, Simulink. This allows the models and controller derived to be tested in a manner representative of a real system. A continuous system is used to prove that fundamentally the system is stable.

The objectives of this section were to prove the aircraft stability under the take-off condition and to show resistance to a constant disturbance. This condition was selected due to it being the most demanding part of the flight, as the engines need to not only power the aircrafts hover, but also provide excess thrust for altitude gain. This test also checks to see if the motor selected has the parameters to provide reactions quick enough to maintain stability. If the motors are too unresponsive, stability will be improbable.

### 4.1.1 Simulation of basic altitude dynamics

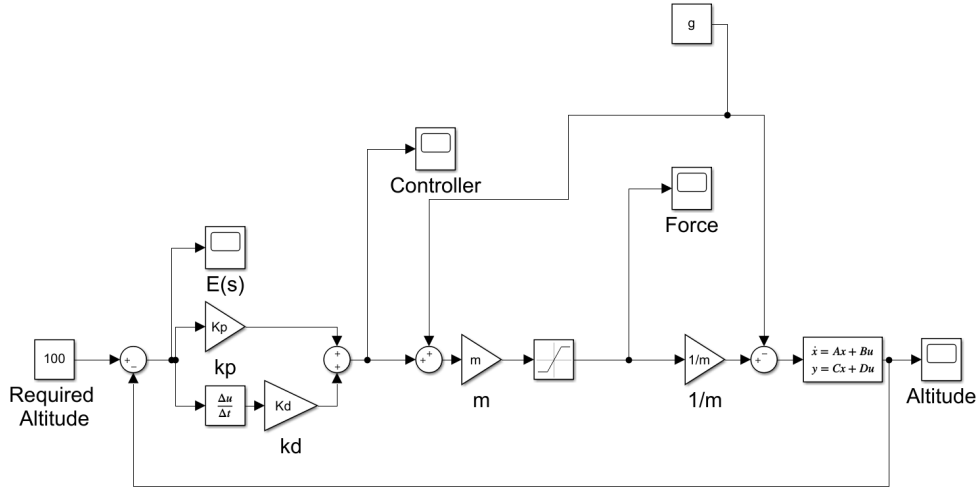

Figure 24 The basic altitude control system

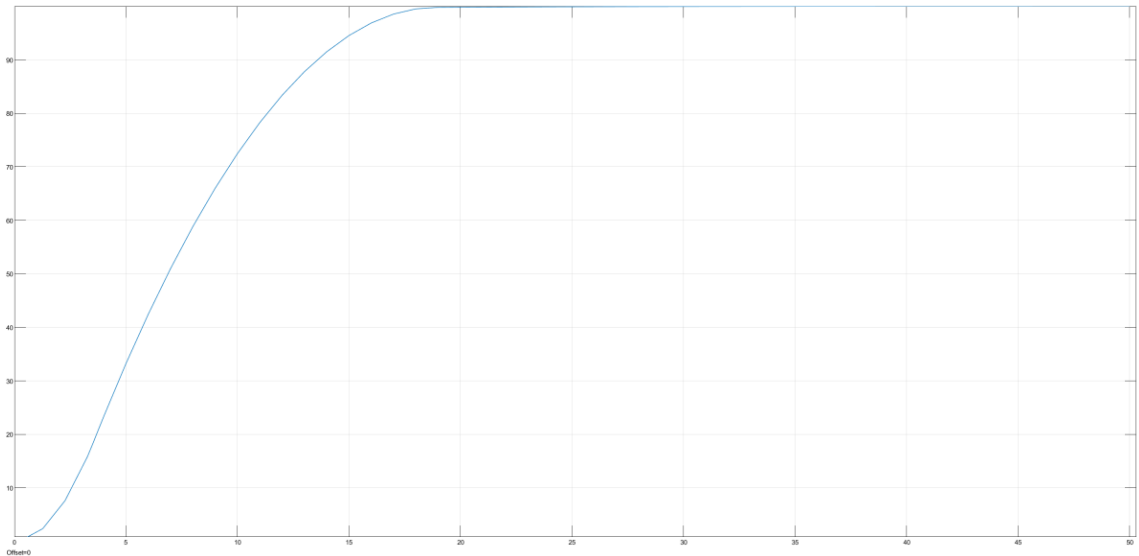

Figures 25 Altitude block diagram and graph from control system in section 4.1.1

Figure 25 is the result of the basic altitude dynamics performed from the control system in Fig. 24. We can see that the craft, even while saturated, can reach a stable altitude in less than 20 seconds. This proves the ability of the aircraft to adjust to altitude commands. This graph assumes the aircraft is flying nominally and is not currently damaged.

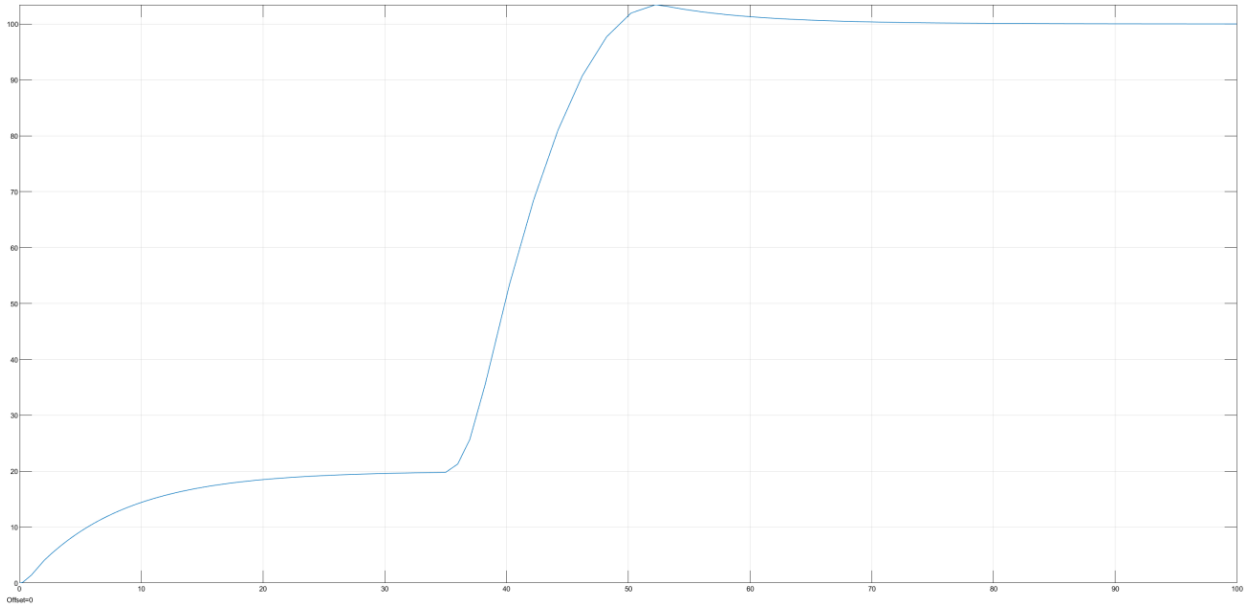

Figure 26 The altitude when the input is a step command

Even when subject to a step input the craft can achieve the required altitudes asked. This is shown above in Fig. 26.

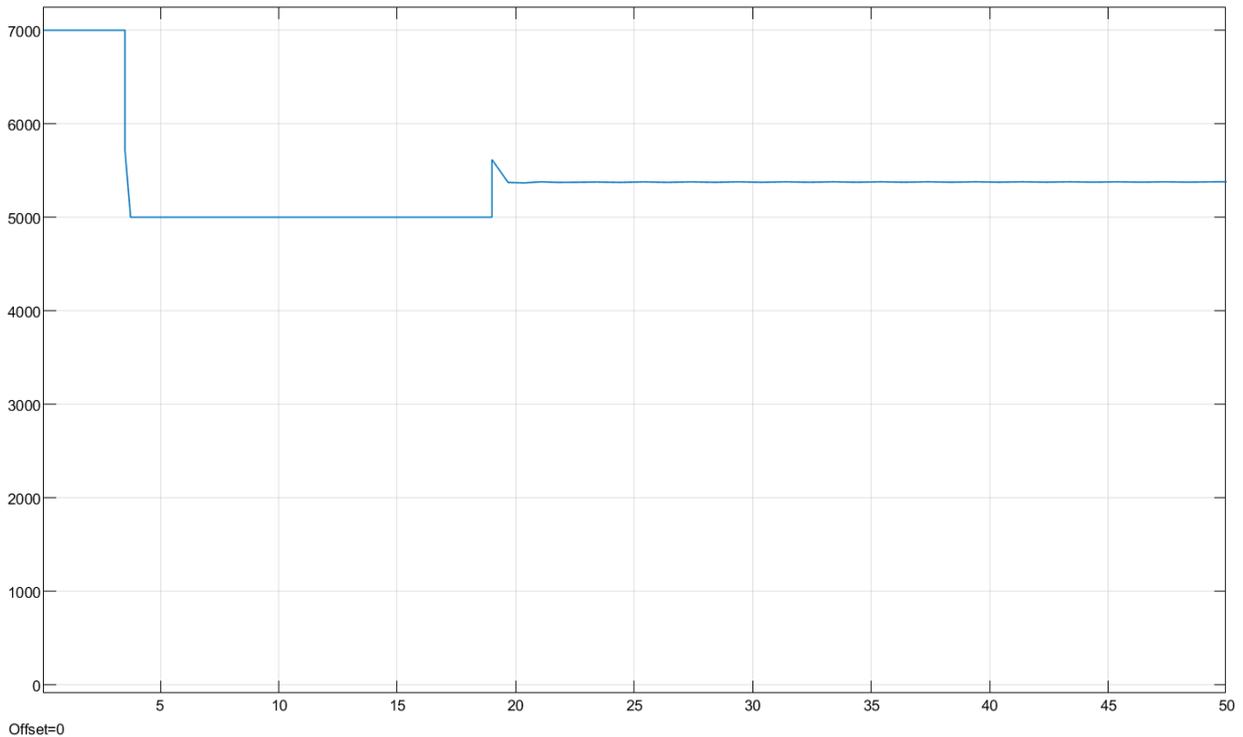

Figure 27 Force plot from control system in section 4.1.1

Figure 27 is the force plot for the altitude graph Fig. 25. We can see the saturation block in effect, where the limit of force is the thrust of the aircraft (729kg thrust or 7100N). This however represents the best-case scenario, where all motors in all ducts are functioning optimally. This however does not describe the behaviour while under the failed state.

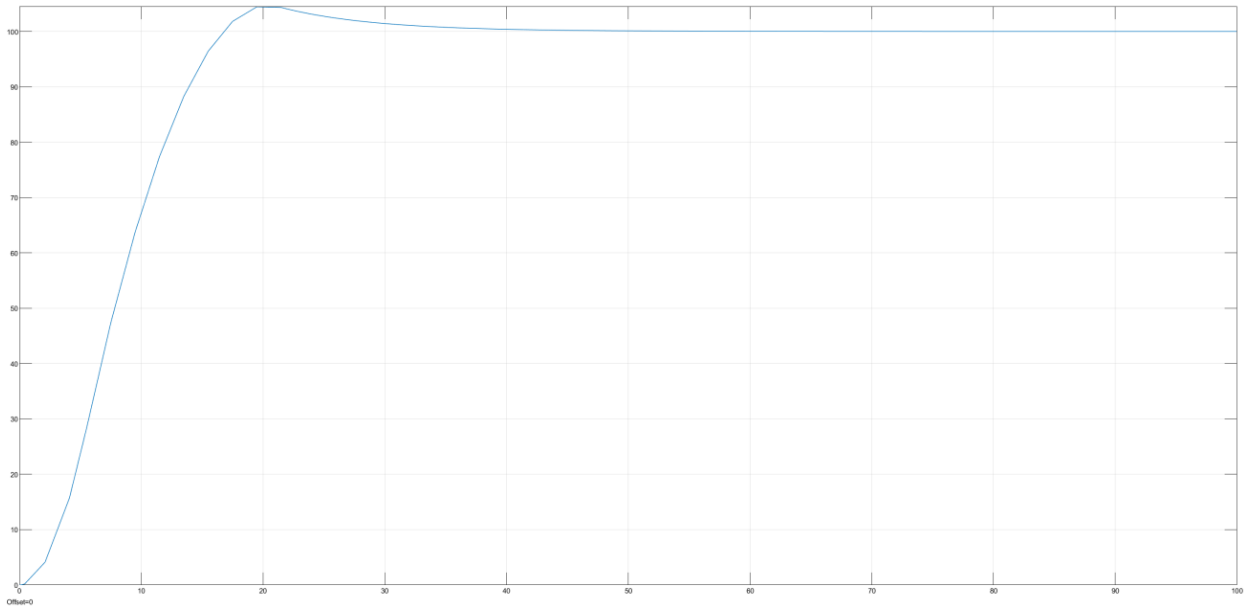

Figure 28 Failed altitude dynamics

When one motor has failed, the aircraft is still able to keep a stable altitude and can continue to fly nominally, as evidenced by Fig. 28. This is not recommended however, as the increased load on the remaining motor is much higher, so the failure chance of the remaining motor is then increased.

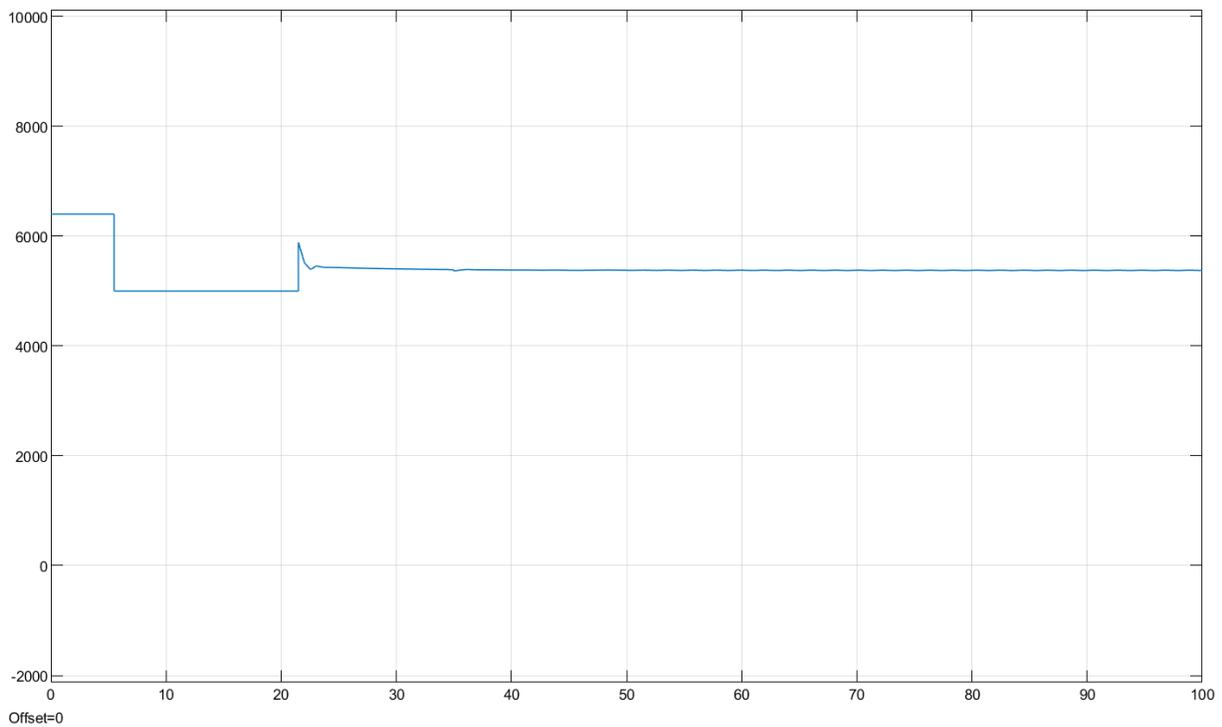

Figure 29 Force plot during failed state (N)

Due to the nature of the aircraft's quadrotor design the force output for a given manoeuvre must remain constant even in a failed state. Therefore, the force plot during a failed state Fig. 29 does not show significantly different data to Fig. 27. However, the remaining motor in the same duct will have a greater output to account for the increased load factor.

### 4.1.2 Simulation of altitude dynamics with motor block

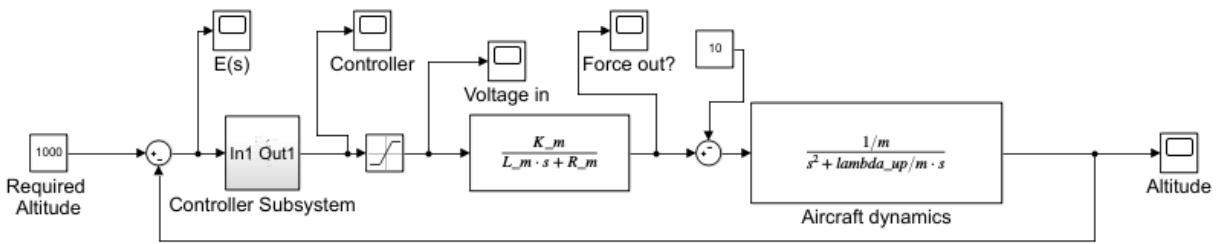

Figure 30 Altitude control system with motor dynamics included

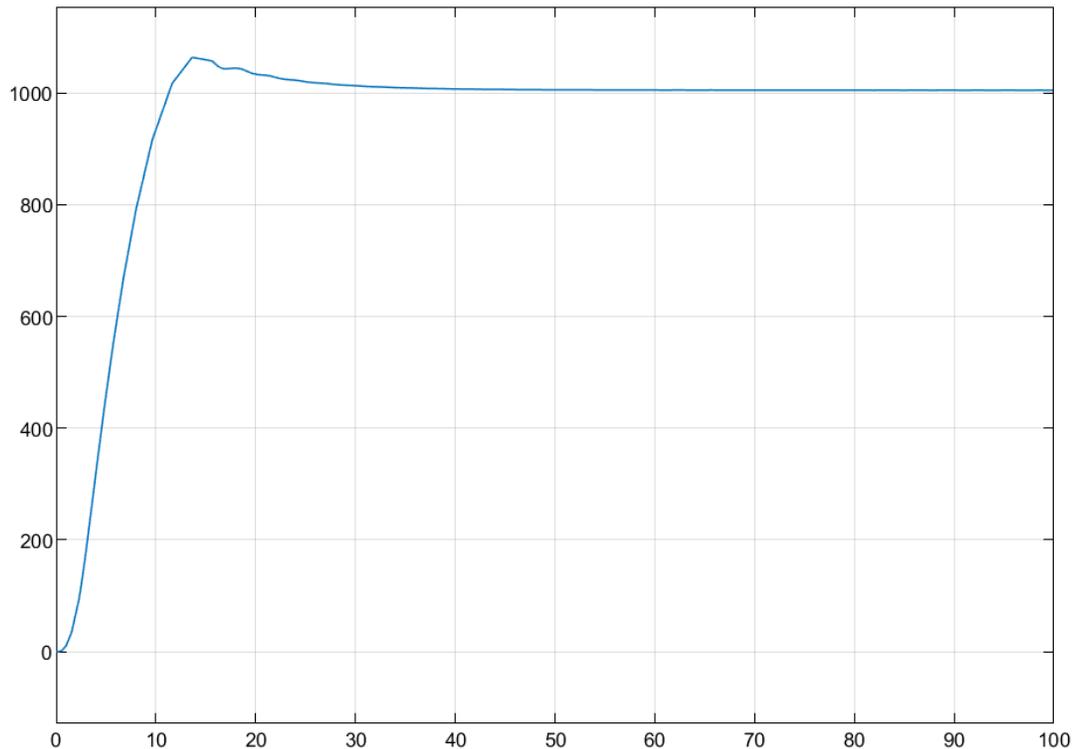

Figures 31 Altitude dynamics with motor included

Figure 30 describes the altitude dynamics with a motor included in the system. We can see that very little change has occurred, and the aircraft is still stable with very little overshoot, shown in Fig. 31. In connection with this, the aircraft can be tuned for greater or less overshoot, depending on rise time requirements.

### 4.1.3 Simulation of roll dynamics with motor block

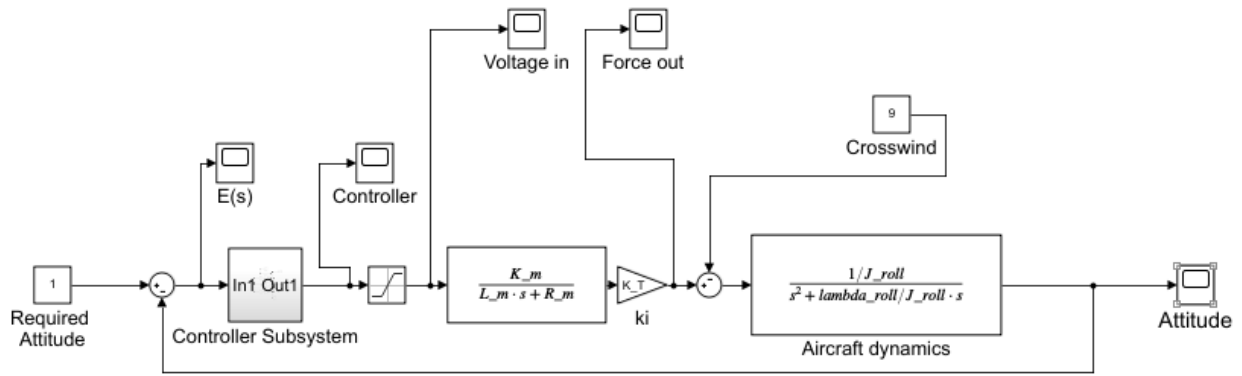

Figure 32 Attitude control system with motor dynamics included

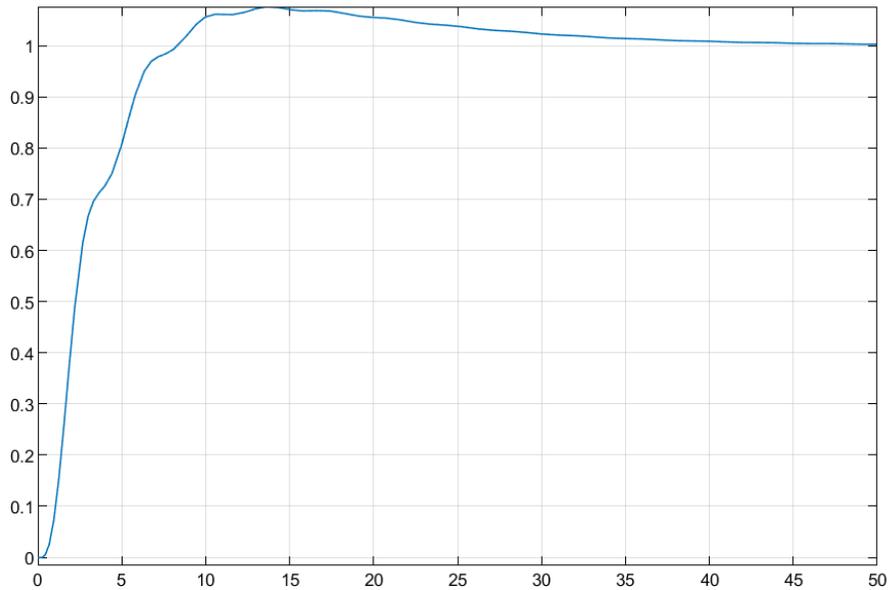

Figures 33 Roll dynamics with error of 1 Rad and heavy crosswind of 9 m/s

Figure 33 shows that aircraft's attitude control system, Fig. 32, can be made stable with very basic tuning. As can be seen, the plot is not smooth. This error can be corrected with further analysis of the roll control parameters. However, the main objective of stable roll control with motor included has been achieved.

## 4.2 Simulation results analysis

After simulating the three most critical conditions, the results were analysed.

1) Analysis for the basic altitude dynamics block

Using the basic altitude system, we can see that the system was able to reach the required altitude in a reasonable time (15 seconds). Even when the input was changed to a step, the system was able to reach the altitude with very little overshoot (2 meters). These results display the systems ability to adapt to new inputs.

2) Analysis for the motor included altitude dynamics block

The results from the motor included altitude dynamics were also promising, showing stable altitude with very little overshoot (4 meters) and a rise time of 11 seconds. This result again shows system stability, and also demonstrates the motors response time is fast enough. Had the motor been slower to respond, the plot would oscillate until failure.

3) Analysis for the motor included attitude dynamics block

Finally the results from the roll dynamics with motor were analysed. It can be seen that the system is stable, however, the plot also displays some pulsing, which is indicative of a parameter mismatched. This pulsing means further refinement of the altitude controller must be done, as this motion is not only a symptom of a bad controller but would also lead to nausea for the pilot.

# 5 CONCLUSION

This paper has been the preliminary design of a large-scale quadrotor. The design was created by collecting previous examples of existing aircraft and choosing design elements which best suited the characteristics required for the design. The aircraft was then formed in design mathematics and simultaneously a 3D model was produced. Finally, a short controllability study was performed. The results lead to a preliminary data sheet, an initial concept 3D model, and proof of stability during the take-off segment. Although more time would have been useful for finalising the 3D model, the events leading up to submission made this impossible.

The aircrafts primary design focus was on safety, and this focus has been achieved. The aircraft can maintain full functionality even in the event than a motor or propeller fails. This ability compared to conventional helicopters is significant given the high failure rate of helicopters, and the complete loss of control when the rotor is damaged. This design focus however lead to the unexpected benefit of providing a large amount of excess thrust, then when directed to the forward direction, could propel the craft at 393 km/h as opposed to the requested top speed of 250 km/h (speed subject to more thorough aerodynamic analysis). Although this speed would not be commonly used, due to the high fuel usage and wear on systems, this ability would be very useful for time critical applications.

The initial design was intended to augment lightweight scout helicopters for military applications. Some secondary design choices include the use of a hybrid powertrain, to increase endurance significantly, and allow ease of refit and refuel, all of which can help increase the viability for the role. The minimum endurance of this craft is 1.5 hours compared to the 25 minuets maximum for existing battery powered air taxis [14]. This addition endurance is the result of the high specific energy of fuel used to power the hybrid system for the majority of flight envelopes. The inclusion of the mass requirements to carry some defensive armaments also parallels existing scout helicopters.

Despite this, the aircraft will also be useful for other applications. These include a high-speed police vehicle which can land vertically in the manner of a conventional helicopter, while also having the range and speed to respond in a much wider area. The armament could be replaced with high resolution imaging equipment for this role. The ability to land without a runway, and travel at high speed allow this aircraft to fill a variety of other roles.

The greatest competitor with the applications of this vehicle are UAVs. UAVs can compete with most the of roles available to an aircraft of this type. They also have the added benefit of not requiring the additional mass of the pilot and systems to inform the pilot. However, the greatest asset of the RQT-3 is the pilot. Many nations impose strict laws against the flying of UAV [19], and so the pilot allows for the aircraft to fill roles otherwise restricted by legislation. Also the pilot can gain greater situational awareness when in controlling the aircraft from the cockpit rather than remotely from a base. Even in the event that restrictions on UAV's become more lenient, the craft could be refitted to be controlled remotely.

## References

[1]  K. T. Waters, "RESEARCH REQUIREMENTS TO IMPROVE SAFETY OF CIVIL HELICOPTERS," NASA, Philadelphia, 1997.

[2]  N. Michael, "Quadrotor Modeling and Control," *Guest Lecture on Aerial Robotics,* pp. 3-13, 2014.

[3]  X. Wang, "Aircraft navigation based on differentiation-integration observer," IIEEE Trans. Ind. Electron., 2017

[4]  A. Chovancová, T. Fico, Ľ. Chovaneca and P. Hubinskýa, "Mathematical Modelling and Parameter Identification of Quadrotor," *ScienceDirect,* pp. 1-10, 2014.

[5]  X. Zhang, X. Li, K. Wang and Y. Lu, "A Survey of Modelling and Identification of Quadrotor Robot," *Abstract and Applied Analysis,* pp. 4-39, 2014.

[6]  B. Douglas, "Drone Simulation and Control," MathWorks, The MathWorks, 2019.

[7]  T. Luukkonen, "Modelling and control of quadcopter," School of Science, Aalto University, 2011.

[8] T. Bresciani, "Modelling, Identification and Control of a Quadrotor Helicopter," Department of Automatic Control, Lund University, 2008.

[9] X. Wang and B. Shirinzadeh, "Nonlinear augmented observer design and application to quadrotor aircraft," *Nonlinear Dynamics,* vol. 80, no. 3, pp. 1463-1481, 2015.

[10] A. Freddi, A. Lanzon and S. Longhi, "A Feedback Linearization Approach to Fault Tolerance in Quadrotor Vehicles," The International Federation of Automatic Control, Milano (Italy), 2011.

[11] J. Ghandour, S. Aberkane and J.-C. Ponsart, "Feedback Linearization approach for Standard and Fault Tolerant control: Application to a Quadrotor UAV Testbed," *Journal of Physics: Conference Series,* pp. 5-25, 2014.

[12] N. P. Nguyen and S. K. Hong, "Fault Diagnosis and Fault-Tolerant Control Scheme for Quadcopter UAVs with a Total Loss of Actuator," *MDPI,* pp. 3-10, 2019.

[13] A. Simha, S. Vadgama and S. Raha, "A Geometric Approach to Rotor Failure Tolerant," *Electrical Engineering,* pp. 4-7, 2017.

[14] EHang, "EHang AAV," 04 December 2019. [Online]. Available: http://www.ehang.com/ehang184.

[15] A. Ahmad, "Speed Control of DC motor using Controllers," researchgate, Dhanbad, 2014.

[16] H. Castaneda, F. Plestan, A. Chriette and J. Leon-Morales, "Continuous Differentiator Based on Adaptive Second-Order Sliding-Mode Control for a 3-DOF Helicopter," IEEE, ICo, 2016.

[17] M. T. A. F. T. M. M. C. P. W. C. E. L. H. D. W. L. R. J. J. M. F. D. F. J. W. P. G. H. Curtis, Aerospace Engineering e-Mega Reference, Florida: Butterworth-Heinemann, 2009.

[18] P. Robertson, "Hybrid Power in Light Aircraft: Design Considerations and Experiences of First Flight," University of Cambridge, Cambridge, 2107.

[19] CAA, "Unmanned aircraft and drones," 26 March 2020. [Online]. Available: https://www.caa.co.uk/Consumers/Unmanned-aircraft-and-drones/.